\documentclass{birkjour}

\usepackage[utf8]{inputenc}		
\usepackage{hyperref}
\usepackage[backend=biber,style=numeric,citestyle=numeric,backref=true,sorting=none,natbib=true,autopunct=true,uniquename=false,uniquelist=false,giveninits]{biblatex}
\DeclareNameAlias{sortname}{last-first}
\usepackage{amssymb}
\usepackage{bm}
\usepackage[usestackEOL]{stackengine}	
\usepackage{scalerel}			

\DeclareMathOperator{\sign}{sign}
\DeclareMathOperator{\Tr}{Tr}
\DeclareMathOperator{\wedgie}{\wedge}

\newcommand{\ee}{e}	
\newcommand{\im}{i}	
\newcommand{\transpose}{\top}

\newcommand{\Cp}{C}			
\newcommand{\conj}[1]{\bar{#1}}		
\newcommand{\reverse}[1]{\tilde{#1}}	
\newcommand\longconj[1]{\ThisStyle{%
  \setbox0=\hbox{$\SavedStyle#1$}%
  \stackengine{1.3\LMpt}{$\SavedStyle#1$}{\rule{\dimexpr\wd0-2\LMpt\relax}{.3\LMpt}}{O}{c}{F}{T}{S}%
}}

\newcommand{\Lchiral}{L}
\newcommand{\Rchiral}{R}
\newcommand{\spinordot}{{\mkern2mu \cdot}}

\newcommand{\ba}{\bm{a}}
\newcommand{\bb}{\bm{b}}

\newcommand{\bepsilon}{\bm{\epsilon}}
\newcommand{\bgamma}{\bm{\gamma}}
\newcommand{\bGamma}{\bm{\Gamma}}

\newcommand{\SO}{{\rm SO}}
\newcommand{\Spin}{{\rm Spin}}
\newcommand{\SU}{{\rm SU}}
\newcommand{\U}{{\rm U}}

\newcommand{\pback}{\mkern-2mu}		

\newcommand\Ttab{\rule{0pt}{2.6ex}}		
\newcommand\Btab{\rule[-1.2ex]{0pt}{0pt}}	

\newcommand{\Upup}{{\Uparrow}{\uparrow}}
\newcommand{\Updown}{{\Uparrow}{\downarrow}}
\newcommand{\Downup}{{\Downarrow}{\uparrow}}
\newcommand{\Downdown}{{\Downarrow}{\downarrow}}

\newcommand{\spinorchartfig}{
    \begin{figure}[t!]
    \begin{center}
    \leavevmode
    \includegraphics[scale=.6]{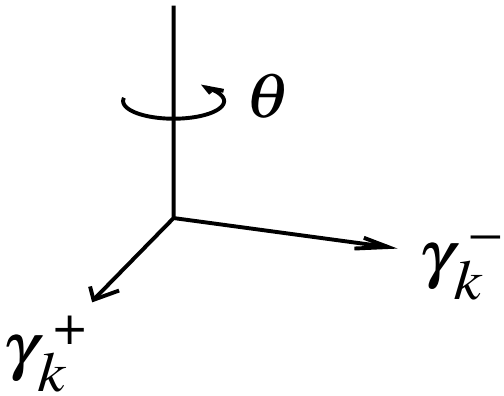}
    \caption[Rotation by angle $\theta$ in $\bgamma^+_k$--$\bgamma^-_k$ plane]{
    \label{spinorchart}
Right-handed rotation by angle $\theta$ in $\bgamma^+_k$--$\bgamma^-_k$ plane.
The $[N/2]$ conserved charges of $\Spin(N)$ are eigenvalues of quantities
under rotations in $[N/2]$ planes
$\bgamma^+_k \bgamma^-_k$, $k = 1 , ... , [N/2]$.
    }
    \end{center}
    \end{figure}
}

\newcommand{\spinormetricsymmetrytab}{
    \begin{table}[t!]
    \caption[Symmetry of spinor metric]{
Symmetry of spinor metric in $N$ spacetime dimensions
    }
    \begin{center}
    \label{spinormetricsymmetrytab}
    \begin{tabular}{ccc}
\hline
\hline
\Ttab
& $\varepsilon^2$ & $\varepsilon_{\rm alt}^2$
\\
$N$ & $(-)^{[(N+1)/4]}$ & $(-)^{[(N+2)/4]}$
\Btab
\\
\hline
1 (mod 8) & $+$ & $+$
\\
2 (mod 8) & $+$ & $-$
\\
3 (mod 8) & $-$ & $-$
\\
4 (mod 8) & $-$ & $-$
\\
5 (mod 8) & $-$ & $-$
\\
6 (mod 8) & $-$ & $+$
\\
7 (mod 8) & $+$ & $+$
\\
8 (mod 8) & $+$ & $+$
\\
\hline
\hline
    \end{tabular}
    \end{center}
    \end{table}
}

\newcommand{\spinormetricgammaatab}{
    \begin{table}[t!]
    \caption[Sign of $\bgamma_a^\transpose \pback \varepsilon = \pm \varepsilon \bgamma_a$]{
Sign of $\bgamma_a^\transpose \pback \varepsilon = \pm \varepsilon \bgamma_a$ in $N$ spacetime dimensions
    }
    \begin{center}
    \label{spinormetricgammaatab}
    \begin{tabular}{ccc}
\hline
\hline
& $\varepsilon$ & $\varepsilon_{\rm alt}$
\\
$N$ & $(-)^{[(N+3)/2]}$ & $(-)^{[N/2]}$
\Btab
\\
\hline
1 (mod 8) & $+$ & $+$
\\
2 (mod 8) & $+$ & $-$
\\
3 (mod 8) & $-$ & $-$
\\
4 (mod 8) & $-$ & $+$
\\
5 (mod 8) & $+$ & $+$
\\
6 (mod 8) & $+$ & $-$
\\
7 (mod 8) & $-$ & $-$
\\
8 (mod 8) & $-$ & $+$
\\
\hline
\hline
    \end{tabular}
    \end{center}
    \end{table}
}

\newcommand{\cpsquaretab}{
    \begin{table}[t!]
    \caption[Symmetry of the conjugation operator]{
Symmetry of the conjugation operator $\Cp$ in $K$+$M$ spacetime dimensions
    }
    \begin{center}
    \label{cpsquaretab}
    \begin{tabular}{ccc}
\hline
\hline
\Ttab
$K - M$ & $\Cp$ & $\Cp_{\rm alt}$
\Btab
\\
\hline
1 (mod 8) & $+$ & $+$
\\
2 (mod 8) & $+$ & $-$
\\
3 (mod 8) & $-$ & $-$
\\
4 (mod 8) & $-$ & $-$
\\
5 (mod 8) & $-$ & $-$
\\
6 (mod 8) & $-$ & $+$
\\
7 (mod 8) & $+$ & $+$
\\
8 (mod 8) & $+$ & $+$
\\
\hline
\hline
    \end{tabular}
    \end{center}
    \end{table}
}

\newcommand{\paritytimetab}{
    \begin{table}[t!]
    \caption[Parity and time reversal operators]{
Parity $P$ and time reversal $T$ operators in $K$+$M$ spacetime dimensions
    }
    \begin{center}
    \label{paritytimetab}
    \begin{tabular}{cccc}
\hline
\hline
\Ttab
$K$ & $M$ & $X = \bgamma_a$ (spacelike) & $X = \bgamma_m$ (timelike)
\Btab
\\
\hline
even & even & $P$ & $T$ \\
odd & odd & $T$ & $P$ \\
even & odd & $PT$ & $-$ \\
odd & even & $-$ & $PT$ \\
\hline
\hline
    \end{tabular}
    \end{center}
    \end{table}
}

\begin{document}

\title[The Supergeometric Algebra]{\boldmath The Supergeometric Algebra}

\author{Andrew J. S. Hamilton}

\address{%
JILA and University of Colorado, Boulder,\\Box 440, Boulder, CO 80309\\USA
}

\email{Andrew.Hamilton@colorado.edu}




\date{\today}

\begin{abstract}
Spinors are central to physics:
all matter (fermions) is made of spinors,
and all forces arise from symmetries of spinors.
It is common to consider the geometric (Clifford) algebra as the fundamental
edifice from which spinors emerge.
This paper advocates the alternative view that spinors are more fundamental
than the geometric algebra.
The algebra consisting of linear combinations of scalars,
column spinors, row spinors, multivectors, and their various products,
can be termed the supergeometric algebra.
The inner product of a row spinor with a column spinor yields a scalar,
while the outer product of a column spinor with a row spinor yields
a multivector, in accordance with the Brauer-Weyl (1935) theorem.
Prohibiting the product of a row spinor with a row spinor,
or a column spinor with a column spinor, reproduces the exclusion principle.
The fact that the index of a spinor is a bitcode is highlighted.
\end{abstract}

\maketitle


\section{Introduction}
\label{sec:intro}


Spinors, objects of spin $\tfrac{1}{2}$, lie at the heart of physics.
On the one hand, all known matter (fermions) is made of spinors.
On the other,
all known forces arise from symmetries of spinors.

Introduced by Cartan \cite{Cartan:1913,Cartan:1938} over a century ago,
a spinor is the fundamental representation of the group $\Spin(N)$
(the covering group of the special orthogonal group $\SO(N)$)
of rotations in $N$ spacetime dimensions.
An intriguing aspect of spinors is that they are indexed by a bitcode,
with $[N/2]$ bits in $N$ spacetime dimensions. 
The halving of dimensions is associated with the natural complex structure
of spinors.
A mathematical physicist would say that the dimension of the spinor
representation of the group $\Spin(N)$
is $2^{[N/2]}$
(technically, in even $N$ dimensions there are two isomorphic irreducible
spinor representations, each of dimension $2^{[N/2]-1}$, of opposite chirality;
chirality counts whether the number of up-bits is even or odd).
Each bit can be either up ($\uparrow$) or down ($\downarrow$).
A spinor in $N = 2$ or 3 dimensions is a Pauli spinor,
with $[N/2] = 1$ bit,
and $2^{[N/2]} = 2$ complex components,
Appendix~\ref{spinorgeom3D-app}.
A spinor in $N = 3{+}1$ spacetime dimensions is a Dirac spinor,
with $[N/2] = 2$ bits,
and $2^{[N/2]} = 4$ complex components,
Appendix~\ref{spinorgeom31D-app}.
The two bits of a Dirac spinor are a spin bit and a boost bit.
The Dirac spinor is said to be right-handed if the spin and boost bits align,
left-handed if they anti-align.

As reviewed by \cite{Baez:2009dj},
the group $\Spin(10)$ of rotations of spinors in 10 dimensions
is a well-known promising candidate
for a grand unified group that contains the standard model group
$\U_Y(1) \times \SU_\Lchiral(2) \times \SU(3)$.
As first pointed out by \cite{Wilczek:1998}
and reviewed by \cite{Baez:2009dj},
$\Spin(10)$ characterizes a generation of fermions with
$[10/2] = 5$ bits $y,z,r,g,b$,
consisting of two weak bits $y$ and $z$,
and three colour bits $r$, $g$, and $b$
(the labelling of bits here follows \cite{Hamilton:2022b}).
Each bit can be either up ($\uparrow$) or down ($\downarrow$),
signifying a charge of $+\tfrac{1}{2}$ or $-\tfrac{1}{2}$.
The charges $y,z,r,g,b$ of each fermion
are linearly related to the 5 traditional conserved charges of the standard
model, which comprise hypercharge $Y$, weak isospin $I_L$, and three colours
\cite{Hamilton:2022b}.

In the $\Spin(10)$ model,
there are $2^{[10/2]} = 2^5 = 32$ fermions in a generation.
Each of the 32 fermions is itself a chiral (massless) fermion,
also called a Weyl fermion,
with 2 complex degrees of freedom,
for a total of $2^6 = 64$ degrees of freedom in a generation.
The components of the Weyl fermion transform under
the group $\Spin(3,1)$ of Lorentz transformations.
Lorentz transformations are of course part of the symmetry group
of general relativity
(the other symmetry of general relativity being translations,
or general coordinate transformations).
The usual assumption, motivated by the Coleman-Mandula no-go theorem
\cite{Coleman:1967,Mandula:2015,Pelc:1997},
is that internal (grand unified) and spacetime groups
combine as a direct product.
However,
the fact that the grand unified group $\Spin(10)$
and the Lorentz group $\Spin(3,1)$ are both spin groups,
and that there are a total of $2^6$ degrees of freedom in a generation
of fermions,
suggests the alternative hypothesis that the groups
might combine non-trivially in a spin group containing $2^6$ spinors,
which, given that there must be a time dimension,
must necessarily be the group $\Spin(11,1)$
in 11+1 spacetime dimensions.

Indeed, \cite{Hamilton:2022b} has shown that the Dirac algebra
(the geometric algebra associated with the Lorentz group $\Spin(3,1)$)
and
the standard model algebra
embed non-trivially as commuting subalgebras of
the $\Spin(11,1)$ geometric algebra,
uniting all four forces of Nature.
The proposed unification in $\Spin(11,1)$
adds one extra bit, the $t$-bit, or time bit,
to the five bits $y,z,r,g,b$ of $\Spin(10)$,
for a total of $[12/2] = 6$ bits.
The unification obeys the Coleman-Mandula theorem
because $\Spin(11,1)$ combines internal and spacetime symmetries
in a single simple group.
After grand symmetry breaking,
the Coleman-Mandula theorem requires only that
{\em unbroken\/} internal symmetries commute with spacetime symmetries;
the unification proposed by
\cite{Hamilton:2022b} achieves just that.
The fact that $\Spin(11,1)$ differs from the Lorentz group $\Spin(3,1)$
by 8 spatial dimensions ensures,
by the period-8 Cartan-Bott
\cite{Cartan:1908,Bott:1970,Coquereaux:1982}
periodicity of geometric algebras,
that the discrete symmetries
(of the spinor metric, Table~\ref{spinormetricsymmetrytab},
and of the conjugation operator, Table~\ref{cpsquaretab})
of the $\Spin(11,1)$ geometric algebra coincide with those of $\Spin(3,1)$.

Besides their behaviour under symmetry transformations,
spinors in physics have various fundamental properties, among which are:
\begin{enumerate}
\item
Spinors possess a (complex-valued) scalar product.
Among other things, the spinor scalar product allows a scalar Lagrangian
for spinors to be defined, the Dirac Lagrangian,
from which the dynamics of spinors flow.
\item
Spinors satisfy an exclusion principle.
\item
Spinors have (complex) conjugates,
which define antispinors.
\end{enumerate}

A major theorem, first proved by Brauer \& Weyl (1935)
\cite{Brauer:1935},
is that the algebra of outer products of spinors
is isomorphic to the Clifford algebra
\cite{Grassmann:1862,Grassmann:1877,Clifford:1878}
of multivectors,
(see \cite{Lounesto:2001} for a history),
also known as the geometric algebra
\cite{Hestenes:1966,Hestenes:1987,Gull:1993,Doran:2003},
in any number of spacetime dimensions.

A dimension-$d$ representation of a group is, by definition,
a mapping from the group to a set of $d \times d$ matrices,
multiplication among which reproduces the action of the group.
The spinors of the spinor representation of the group $\Spin(N)$
are column vectors of dimension $2^{[N/2]}$.
In physics there are also row spinors,
a product of row and column spinors yielding a scalar product of spinors.
Elements of the rotation group $\Spin(N)$ acting on spinors are called rotors,
represented as $2^{[N/2]} \times 2^{[N/2]}$ matrices.
Rotors comprise the set of even multivectors
obtained by exponentiating the orthonormal bivectors of the geometric algebra.

One of the core properties of spinors in physics
is that they satisfy an exclusion principle.
The exclusion principle underlies much of the richness
of the behaviour of matter at low energy.
According to the usual rules of matrix multiplication,
a row matrix can multiply a column matrix,
yielding a scalar,
and
a column matrix can multiply a row matrix,
yielding a matrix,
but a row matrix cannot multiply a row matrix,
and
a column matrix cannot multiply a column matrix:
\begin{subequations}
\label{spinormultiplication}
\begin{alignat}{3}
  \begin{array}{@{}c@{}}
  \begin{pmatrix} \phantom{0} & \phantom{0} & \phantom{0} \end{pmatrix}
  \\ \phantom{0} \\ \phantom{0}
  \end{array}
  \scalebox{.67}[1]{$\begin{pmatrix} \phantom{M} \\ \phantom{M} \\ \phantom{M} \end{pmatrix}$}
  &
  \begin{matrix} \ =\, \\ \phantom{0} \\ \phantom{0} \end{matrix}
  \begin{array}{@{}c@{}}
  \begin{pmatrix} \phantom{0} \end{pmatrix}
  \\ \phantom{0} \\ \phantom{0}
  \end{array}
  &&
  \begin{array}{@{}c@{}}
  \quad
  \mbox{inner product = scalar} \ ,
  \\ \phantom{0} \\ \phantom{0}
  \end{array}
\\
  \scalebox{.67}[1]{$\begin{pmatrix} \phantom{M} \\ \phantom{M} \\ \phantom{M} \end{pmatrix}$}
  \begin{array}{@{}c@{}}
  \begin{pmatrix} \phantom{0} & \phantom{0} & \phantom{0} \end{pmatrix}
  \\ \phantom{0} \\ \phantom{0}
  \end{array}
  &
  \begin{matrix} \ =\, \\ \phantom{0} \\ \phantom{0} \end{matrix}
  \scalebox{.67}[1]{$\begin{pmatrix} \phantom{M} & \phantom{M} & \phantom{M} \\ \phantom{M} & \phantom{M} & \phantom{M} \\ \phantom{M} & \phantom{M} & \phantom{M} \end{pmatrix}$}
  &&
  \begin{array}{@{}c@{}}
  \quad
  \mbox{outer product = multivector} \ ,
  \\ \phantom{0} \\ \phantom{0}
  \end{array}
\\
  \begin{pmatrix} \phantom{0} & \phantom{0} & \phantom{0} \end{pmatrix}
  \begin{pmatrix} \phantom{0} & \phantom{0} & \phantom{0} \end{pmatrix}
  &
  \begin{matrix} \ =\, \end{matrix}
  \begin{matrix} \ \varnothing \end{matrix}
  &&
  \quad
  \mbox{forbidden} \ ,
\\
  \scalebox{.67}[1]{$\begin{pmatrix} \phantom{M} \\ \phantom{M} \\ \phantom{M} \end{pmatrix}$}
  \scalebox{.67}[1]{$\begin{pmatrix} \phantom{M} \\ \phantom{M} \\ \phantom{M} \end{pmatrix}$}
  &
  \begin{matrix} \ =\, \\ \phantom{0} \\ \phantom{0} \end{matrix}
  \begin{matrix} \ \varnothing \\ \phantom{0} \\ \phantom{0} \end{matrix}
  &&
  \begin{array}{@{}c@{}}
  \quad
  \mbox{forbidden} \ .
  \\ \phantom{0} \\ \phantom{0}
  \end{array}
\end{alignat}
\end{subequations}
These rules resemble the rules for fermionic creation and destruction operators
in quantum field theory:
creation following destruction is allowed,
and
destruction following creation is allowed,
but
creation following creation is forbidden,
and
destruction following destruction is forbidden.
It is beyond the scope of this paper to pursue the subject,
but it can be shown that
the multiplication rules for row and column spinors indeed
reproduce those of fermion creation (row) and destruction (column)
operators in quantum field theory.

The observed properties of spinors in physics,
as summarized above,
motivate a principal purpose of this paper,
which is to introduce the supergeometric algebra,
defined to consist of (complex) linear combinations
of scalars (= complex numbers), column spinors, row spinors, and multivectors,
and the various multiplication rules that connect them.
The name supergeometric algebra
accords with the common physics practice of prepending the adjective
`super' to spinorial extensions of theories.

The present paper sets out the notation and groundwork for
the subsequent paper \cite{Hamilton:2022b}.
From a purely mathematical perspective,
most of the content of the present paper can be considered well known.
However, the perspective and emphasis are different.
The exposition is intended to be self-contained.


This paper adopts a trailing dot notation for row spinors,
equation~(\ref{rowspinor}).
The notation $\psi \spinordot$ for a row spinor,
with a trailing dot symbolizing the spinor metric,
is extremely convenient.
The dot immediately distinguishes a row spinor from a column spinor;
and the dot makes transparent the application of the associative rule
to a sequence of products of spinors, equation~(\ref{spinorassociative}).

The Brauer-Weyl theorem \cite{Brauer:1935} has practical implications.
A consequence of the theorem is that manipulations commonly carried out
using the geometric algebra can also be carried out using
the supergeometric algebra.
For example,
amplitudes of processes involving spinors in quantum field theory
are commonly calculated using ``$\gamma$-matrix technology,''
also known as Fierz identities, e.g.\ \cite{Peskin:1995}.
Those calculations can be done more directly using the supergeometric algebra
and the associative rule~(\ref{spinorassociative}).
For the same reason that it is simpler to square a quantity than to take
its square root, it is simpler to go from the supergeometric algebra
to the geometric algebra than vice versa.

Two of the most fertile ideas of modern physics are
supersymmetry, and string theory,
although as yet neither has received any experimental corroboration.
Supersymmetry hypothesizes a local gauge theory with spinor generators
\cite{vanNieuwenhuizen:1981,vanNieuwenhuizen:2004rh,Ferrara:2017hed}.
An exciting implication of supersymmetry
is that a spacetime translation is generated by a product of
spinor transformations.
Supersymmetry is built on the relation between spinors and vectors
embodied by the part of the supergeometric algebra
for which the outer products of spinors yield vectors, multivectors of grade~1.
In 4 spacetime dimensions, the relation between spinors and vectors
is given by equations~(\ref{quadrupletdirac}).

But the supergeometric algebra contains not just vectors,
but multivectors of all grades.
A multivector of grade $p$, a $p$-vector,
is a $p$-dimensional object in $N$-dimensional spacetime.
Such things are called branes in string theory.
The algebra of outer products of spinors yields branes
of all dimensions.
The supergeometric algebra would seem to supply a natural language
for string theory.

The plan of this paper is as follows.
Section~\ref{multivectorspinor-sec}
defines the properties of multivectors and spinors.
Section~\ref{algebraouterproduct-sec}
proves the Brauer-Weyl \cite{Brauer:1935} theorem
in even spacetime dimensions $N$.
Section~\ref{spinalgebraodd-sec} addresses the case of odd dimensions $N$.
Section~\ref{chiral-sec}
comments briefly on chiral subalgebras.
Section~\ref{conjugation-sec}
discusses the operation of conjugation,
which translates spinors into their complex conjugates, antispinors,
in a Lorentz-invariant fashion.
The term ``Lorentz-invariant'' is used throughout this paper
to denote invariance under rotations
in an arbitrary number of spacetime dimensions,
not just 3+1 dimensions.
Section~\ref{conclusions-sec}
summarizes the conclusions.


\section{Multivectors and spinors}
\label{multivectorspinor-sec}

\subsection{Multivectors}

The orthonormal vectors $\bgamma_a$, $a = 1 , ... , N$
of a geometric algebra (Clifford algebra)
in $N$ spacelike dimensions satisfy the product rule
\begin{equation}
  \bgamma_a \bgamma_b
  =
  \left\{
  \begin{array}{ll}
  \delta_{ab} & ( a = b ) \ ,
  \\
  - \bgamma_b \bgamma_a & ( a \neq b ) \ ,
  \end{array}
  \right.
\end{equation}
where the Kronecker delta $\delta_{ab}$ is the Euclidean metric.
For the most part,
a dimension that is timelike can be treated as the imaginary $\im$
times a spacelike vector;
the only place where that algorithm is insufficient is conjugation,
section~\ref{conjugation-sec}.
An antisymmetric product of vectors is commonly written with a wedge sign,
$\bgamma_a \wedgie \bgamma_b$,
although in the present paper the wedge sign is often omitted for brevity
when there is no ambiguity.
A wedge product
\begin{equation}
  \bgamma_A
  =
  \bgamma_{a_1 ... a_p}
  \equiv
  \bgamma_{a_1} \wedgie ... \wedgie \bgamma_{a_p}
\end{equation}
of $p$ vectors, a multivector of grade $p$, or $p$-vector,
represents a $p$-dimensional element in the $N$-dimensional space.
The geometric algebra is the $2^N$-dimen\-sional algebra
of linear combinations
\begin{equation}
  \ba
  =
  a^A \bgamma_A
  \ ,
\end{equation}
of multivectors of all possible grades $p = 0$ to $N$.

The group $\Spin(N)$ of rotations in $N$ dimensions
is generated by the $N(N{-}1)/2$ bivectors of the geometric algebra.
Of these bivectors, $[N/2]$ are mutually commuting,
and there are $[N/2]$ associated conserved charges,
called central charges.
Each of the mutually commuting bivectors generates a rotation
in a plane defined by a pair of dimensions.
The $N$ orthonormal basis vectors $\bgamma_a$ thus organize into $[N/2]$ pairs,
conveniently denoted $\bgamma_k^\pm$.
If the dimension $N$ is odd,
then one basis vector, $\bgamma_N$, remains unpaired.
Each of the $[N/2]$ indices $k$ represents a conserved charge.
Chiral combinations
$\bgamma_k$ and $\bgamma_{\bar  k}$
of the orthonormal basis vectors are defined by
\begin{equation}
\label{gammaapm}
  \bgamma_k
  \equiv
  {\bgamma_k^+ + \im \bgamma_k^- \over \sqrt{2}}
  \ , \quad
  \bgamma_{\bar  k}
  \equiv
  {\bgamma_k^+ - \im \bgamma_k^- \over \sqrt{2}}
  \ ,
\end{equation}
which are normalized so that $\bgamma_k \cdot \bgamma_{\bar  k} = 1$.
The vectors
$\bgamma_k^+$ and $\im \bgamma_k^-$
can be thought of as, modulo a normalization,
the real and imaginary parts of a complex vector
$\bgamma_k$ whose complex conjugate is $\bgamma_{\bar  k}$.
Under a right-handed rotation by angle $\theta$ in the
$\bgamma_k^+ \bgamma_k^-$ plane,
Figure~\ref{spinorchart},
the chiral basis vectors $\bgamma_k$ and $\bgamma_{\bar  k}$
transform by a phase,
\begin{equation}
\label{gammaNplusminustransform}
  \bgamma_k
  \rightarrow
  \ee^{- \im \theta} \,
  \bgamma_k
  \ , \quad
  \bgamma_{\bar  k}
  \rightarrow
  \ee^{\im \theta} \,
  \bgamma_{\bar  k}
  \ .
\end{equation}
(If one of the two orthonormal dimensions, say $\bgamma^-_k$,
is a time dimension, then
the rotation in the $\bgamma_k^+ \bgamma_k^-$ plane
becomes a Lorentz boost,
and the transformation~(\ref{gammaNplusminustransform}) becomes
\begin{equation}
\label{gammaNplusminusboost}
  \bgamma_k
  \rightarrow
  \ee^{\theta} \,
  \bgamma_k
  \ , \quad
  \bgamma_{\bar k}
  \rightarrow
  \ee^{- \theta} \,
  \bgamma_{\bar k}
  \ . )
\end{equation}
The transformation~(\ref{gammaNplusminustransform})
identifies the chiral basis vectors $\bgamma_k$ and $\bgamma_{\bar  k}$
as having $k$-charge equal to $+1$ and $-1$.
All other chiral basis vectors,
$\bgamma_l$ and $\bgamma_{\bar l}$ with $l \neq k$,
along with the unpaired basis vector $\bgamma_N$ if $N$ is odd,
remain unchanged under a rotation in the
$\bgamma_k^+ \bgamma_k^-$ plane,
so have zero $k$-charge.
The $k$-charge of a multivector (or tensor of multivectors) can be read off
from its covariant chiral indices:
\begin{equation}
\label{kcharge}
  \mbox{$k$-charge}
  =
  \mbox{number of $k$ minus ${\bar  k}$ covariant chiral indices}
  \ .
\end{equation}

\spinorchartfig

For ``ordinary'' spatial rotations,
the conserved charge is the projection of the angular momentum, or spin,
in the $\bgamma_k^+ \bgamma_k^-$ plane (in fundamental units, $\hbar = 1$).
However,
in other applications of the supergeometric algebra,
the word charge may refer to other conserved charges,
such as the conserved charges of the standard model of physics
\cite{Hamilton:2022b}.

\subsection{Spinors}
\label{basisspinors-sec}

Chiral basis spinors comprise $2^{[N/2]}$ basis spinors $\bepsilon_a$,
\begin{equation}
\label{spinoraxesN}
  \bepsilon_a
  \equiv
  \bepsilon_{a_1 ... a_{[N/2]}}
\end{equation}
where $a = a_1 ... a_{[N/2]}$ denotes a bitcode of length $[N/2]$.
Each bit $a_k$ is either up $\uparrow$ or down $\downarrow$.
For example,
one of the basis spinors is the all-bit-up basis spinor
$\bepsilon_{\uparrow\uparrow ... \uparrow}$.
A spinor $\psi$,
\begin{equation}
  \psi = \psi^a \bepsilon_a
  \ ,
\end{equation}
is a
linear combination of the $2^{[N/2]}$ basis spinors $\bepsilon_a$.

Under a right-handed rotation by angle $\theta$ in the
$\bgamma_k^+ \bgamma_k^-$ plane,
basis spinors
$\bepsilon_{... k ...}$
and
$\bepsilon_{... {\bar  k} ...}$
with $k$-bit respectively up and down
transform as
\begin{equation}
\label{gammaNupdowntransform}
  \bepsilon_{... k ...}
  \rightarrow
  \ee^{- \im \theta / 2} \,
  \bepsilon_{... k ...}
  \ , \quad
  \bepsilon_{... {\bar  k} ...}
  \rightarrow
  \ee^{\im \theta / 2} \,
  \bepsilon_{... {\bar  k} ...}
  \ .
\end{equation}
(Again, if one of the two orthonormal dimensions, say $\bgamma^-_k$,
is a time dimension, then
the rotation in the $\bgamma_k^+ \bgamma_k^-$ plane
becomes a Lorentz boost,
and the transformation~(\ref{gammaNupdowntransform}) becomes
\begin{equation}
\label{gammaNupdownboost}
  \bepsilon_{... k ...}
  \rightarrow
  \ee^{\theta / 2} \,
  \bepsilon_{... k ...}
  \ , \quad
  \bepsilon_{... {\bar  k} ...}
  \rightarrow
  \ee^{- \theta / 2} \,
  \bepsilon_{... {\bar  k} ...}
  \ . )
\end{equation}
The transformation~(\ref{gammaNupdowntransform})
shows that basis spinors
$\bepsilon_{... k ...}$
and
$\bepsilon_{... {\bar  k} ...}$
have $k$-charge respectively
$+\frac{1}{2}$ and $-\tfrac{1}{2}$
in each of its $[N/2]$ bits.
The $k$-charge of a spinor (or tensor of spinors) can be read off
from its covariant chiral indices:
\begin{equation}
\label{kchargespinor}
  \mbox{$k$-charge}
  =
  \tfrac{1}{2}
  ( \mbox{number of $k$ minus ${\bar  k}$ covariant chiral indices} )
  \ .
\end{equation}

\subsection{Rotors}

A rotation in the geometric algebra is described by a rotor $R$,
an element of the rotor group $\Spin(N)$,
the Lie group generated by the Lie algebra of bivectors.
A multivector $\ba$ transforms under a rotor $R$ as
\begin{equation}
\label{rotormultivector}
  R : \ 
  \ba \rightarrow R \ba \reverse{R}
  \ ,
\end{equation}
where $\reverse{R} = R^{-1}$ is the reverse of $R$.
A spinor $\psi$ transforms under a rotor $R$ as
\begin{equation}
\label{rotorspinor}
  R : \ 
  \psi \rightarrow R \psi
  \ .
\end{equation}
A rotor $R$ is an even, unimodular ($R \reverse{R} = 1$) multivector,
but only in dimensions $N \leq 5$ are all even, unimodular multivectors
also rotors.

The action of the $N ( N{-}1 ) / 2$ bivectors $S_A$
on the $2^{[N/2]}$ basis spinors $\bepsilon_a$
defines a set of $2^{[N/2]} \times 2^{[N/2]}$ matrices,
\begin{equation}
  S_A : \
  \bepsilon_a \rightarrow ( S_A )_{ab} \bepsilon_b
  \ .
\end{equation}
The spinors $\bepsilon_a$ and the accompanying set of matrices $( S_A )_{ab}$
define the $2^{[N/2]}$ dimensional spinor representation of the rotor group.
The matrices of the spinor representation depend on the choice of basis spinors.
Section~\ref{chiralrep-sec} derives the spinor representation
in the chiral representation,
that is, with respect to the chiral basis of spinors.

\subsection{Spinor metric}
\label{spinormetric-sec}

The spinor metric $\varepsilon$ is as fundamental to the supergeometric algebra
as the Euclidean
(or Minkowski)
metric is to the geometric algebra.
The spinor metric $\varepsilon$ is that spinor tensor that is
invariant under rotations, suitably normalized,
in the same way that the Euclidean
metric is that vector tensor that is
invariant under rotations, suitably normalized.
Cartan \cite{Cartan:1938}
calls the spinor metric
the fundamental polar, which he denotes $C$.
It is worth remarking that
if all dimensions are spacelike,
them the spinor metric $\varepsilon$
coincides with the conjugation operator, here denoted $\Cp$,
section~\ref{conjugationop-sec};
but if one or more dimensions are timelike,
then the spinor metric $\varepsilon$ and conjugation operator $\Cp$ differ,
equation~(\ref{arbdimensionCp}).

A scalar product $\psi \cdot \chi$ of spinors $\psi$ and $\chi$ transforms
under a rotor $R$ as
\begin{equation}
\label{rotorscalarproduct}
  R : \ 
  \psi \cdot \chi
  \equiv
  \psi^\transpose \pback \varepsilon \chi
  \rightarrow
  \psi^\transpose \pback R^\transpose \pback \varepsilon R \chi
  =
  \psi^\transpose \pback \varepsilon \chi
  \ ,
\end{equation}
so the spinor metric must satisfy
$R^\transpose \pback \varepsilon R = \varepsilon$,
or equivalently
\begin{equation}
\label{pspinormetrictensorN}
  R^\transpose \pback \varepsilon
  =
  \varepsilon \reverse{R}
  \ .
\end{equation}
The condition~(\ref{pspinormetrictensorN})
is determined by the commutation properties of the spinor metric $\varepsilon$
with the orthonormal bivectors of the geometric algebra.
Condition~(\ref{pspinormetrictensorN}) holds
if $\varepsilon$ commutes with orthonormal basis bivectors
whose representation is real,
and anticommutes with orthonormal basis bivectors
whose representation is imaginary
(in this context, a timelike vector is to be construed
as $\im$ times a spacelike orthonormal vector).
In the chiral representation
constructed in section~\ref{chiralrep-sec},
all chiral basis vectors
$\bgamma_k$ and $\bgamma_{\bar  k}$
are real,
so in the chiral representation orthonormal basis vectors $\bgamma_k^+$
(along with $\bgamma_N$ if $N$ is odd) are real,
while $\bgamma_k^-$ are imaginary,
equations~(\ref{gammaapm}).
The only matrix $\varepsilon$ with the required commutation properties
with basis bivectors is,
up to a scalar or pseudoscalar normalization factor,
the product of the basis vectors $\bgamma_k^+$
(along with $\bgamma_N$ if $N$ is odd),
\begin{equation}
\label{pspinormetricp}
  \varepsilon
  =
  \prod_{k=1}^{[(N+1)/2]} \bgamma_k^+
  \ ,
\end{equation}
where the final factor $\bgamma_{[(N+1)/2]}^+$
is to be interpreted as $\bgamma_N$ if $N$ is odd.
An alternative version
$\varepsilon_{\rm alt}$ of the spinor metric
may be obtained by multiplying the spinor metric~(\ref{pspinormetricp})
by the chiral factor $\varkappa_N$,
which is the pseudoscalar (product of all $N$ vectors)
normalized by a power of $\im$ so that $\varkappa_N^2$ equals one,
equation~(\ref{kappaN}),
\begin{equation}
\label{altpspinormetricp}
  \varepsilon_{\rm alt}
  \equiv
  \varkappa_N
  \varepsilon
  =
  \prod_{k=1}^{[N/2]} \im \bgamma_k^-
  \ .
\end{equation}
The factors of the imaginary $\im$
are introduced so that the spinor metric $\varepsilon_{\rm alt}$ is real.
Notwithstanding the equality of $\varepsilon$ and $\prod_k \bgamma_k^+$
(or of $\varepsilon_{\rm alt}$ and $\prod_k \im \bgamma_k^-$)
in the chiral representation,
$\varepsilon$ (or $\varepsilon_{\rm alt}$)
is defined to transform as a Lorentz-invariant spinor tensor under rotations,
not as an element of the geometric algebra.
The vectors $\bgamma_k^+$ and $\bgamma_k^-$
in the expressions~(\ref{pspinormetricp}) and~(\ref{altpspinormetricp})
are to be interpreted as being spacelike
(timelike vectors being treated as $\im$ times a spacelike vector).
The spinor metric $\varepsilon$ or $\varepsilon_{\rm alt}$ is the same
independent of whether orthonormal vectors are spacelike or timelike.

If $N$ is odd,
and if the odd algebra is constructed
as described in section~\ref{oddN2-sec},
then there are potentially further options for the spinor metric,
relegated to Appendix~\ref{spinormetric-app}.

\spinormetricsymmetrytab

The spinor metric $\varepsilon$,
in either of the forms~(\ref{pspinormetricp}) or~(\ref{altpspinormetricp}),
is real and orthogonal,
and its square is plus or minus the unit matrix,
\begin{equation}
\label{pspinormetricsquared}
  \varepsilon^{-1} = \varepsilon^\transpose
  \ , \quad
  \varepsilon^2
  =
  \pm 1
  \ ,
\end{equation}
where the $\pm$ sign is as tabulated in
Table~\ref{spinormetricsymmetrytab}.
For odd $N$,
the sign agrees with that of Cartan
\cite{Cartan:1938}, eq.~(13), \S101 on p.~88;
for even $N$,
the sign agrees with Cartan's
for the alternative spinor metric $\varepsilon_{\rm alt}$,
and is minus Cartan's for the standard spinor metric $\varepsilon$.
Table~\ref{spinormetricsymmetrytab}
exhibits the well-known period-8 Cartan-Bott
\cite{Cartan:1908,Bott:1970,Coquereaux:1982}
periodicity of geometric algebras.
For Dirac spinors in $N = 4$ spacetime dimensions,
the spinor metric is the standard spinor metric~(\ref{pspinormetricp}),
which is antisymmetric according to Table~\ref{spinormetricsymmetrytab}.

According to equation~(\ref{pspinormetricNN2})
(or~(\ref{altpspinormetricNN2}) for $\varepsilon_{\rm alt}$),
the components $\varepsilon_{ba}$
of the spinor metric $\varepsilon$,
\begin{equation}
  \varepsilon_{ba}
  \equiv
  \bepsilon_b \cdot \bepsilon_a
  \equiv
  \bepsilon_b^\transpose \varepsilon \bepsilon_a
  \ ,
\end{equation}
are non-vanishing only between basis spinors
$\bepsilon_b$ and $\bepsilon_a$
that are bit flips of each other,
as must be so because the scalar product must have zero charge.
The sign of $\varepsilon_{\bar{a}a}$,
where $\bar{a}$ denotes the bit flip of $a$,
follows inductively from equations~(\ref{bepsilonrep}),
and is
\begin{equation}
\label{signpspinormetric}
  \varepsilon \bepsilon_a
  =
  \sign ( \varepsilon_{\bar{a}a} )
  \bepsilon_{\bar a}
  \ , \quad
  \sign ( \varepsilon_{\bar{a}a} )
  \equiv
  \sign ( \varepsilon_{\bar{a}_1 ... \bar{a}_{[N/2]} a_1 ... a_{[N/2]}} )
  =
  \prod_{a_k = \,\uparrow} (-)^{k-1}
  \ .
\end{equation}
For the alternative spinor metric~(\ref{altpspinormetricp}),
the sign is
\begin{equation}
\label{signaltpspinormetric}
  \varepsilon_{\rm alt} \bepsilon_a
  =
  \sign ( \varepsilon^{\rm alt}_{\bar{a}a} )
  \bepsilon_{\bar a}
  \ , \quad
  \sign ( \varepsilon^{\rm alt}_{\bar{a}a} )
  \equiv
  \sign ( \varepsilon^{\rm alt}_{\bar{a}_1 ... \bar{a}_{[N/2]} a_1 ... a_{[N/2]}} )
  =
  \prod_{a_k = \,\uparrow} (-)^{k}
  \ .
\end{equation}
The spinor metric is symmetric or antisymmetric
as its square is positive or negative,
\begin{equation}
\label{symmetrypspinormetric}
  \varepsilon_{ab}
  =
  \pm \varepsilon_{ba}
  \ ,
\end{equation}
where the $\pm$ sign is as tabulated in Table~\ref{spinormetricsymmetrytab}.

\spinormetricgammaatab

Commuting the spinor metric $\varepsilon$ through
the orthonormal basis vectors $\bgamma_a$
converts them to plus or minus their transposes,
\begin{equation}
\label{pspinormetriccommutep}
  \bgamma_a^\transpose \pback \varepsilon
  =
  \pm
  \varepsilon
  \bgamma_a
  \ .
\end{equation}
Table~\ref{spinormetricgammaatab}
tabulates the sign in equation~(\ref{pspinormetriccommutep})
for the spinor metric $\varepsilon$
and the alternative spinor metric $\varepsilon_{\rm alt}$.

Equation~(\ref{pspinormetriccommutep}) implies that
the commutation rule of an orthonormal multivector $\bgamma_A$ of grade $p$
with the spinor metric $\varepsilon$ is
\begin{equation}
\label{pspinormetriccommutepp}
  \bgamma_A^\transpose \varepsilon
  =
  (\pm)^p
  \varepsilon
  \reverse{\bgamma}_A
  =
  (\pm)^p (-)^{[p/2]}
  \varepsilon
  \bgamma_A
  \ ,
\end{equation}
where $\reverse{\bgamma}_A$ is the reverse of $\bgamma_A$,
and the $\pm$ sign in $(\pm)^p$ is that
in equation~(\ref{pspinormetriccommutep}), which depends on dimension $N$
as tabulated in Table~\ref{spinormetricgammaatab}.

Indices on a spinor are raised and lowered by pre-multiplying by the
spinor metric and its inverse,
\begin{equation}
  \bepsilon^a
  =
  \varepsilon^{ab}
  \bepsilon_b
  \ , \quad
  \bepsilon_b
  =
  \varepsilon_{ba}
  \bepsilon^a
  \ .
\end{equation}

\subsection{Column and row spinors}
\label{columnrowspinor-sec}

Corresponding to any column basis spinor $\bepsilon_a$
is a row basis spinor $\bepsilon_a \spinordot$ defined by
\begin{equation}
  \bepsilon_a \spinordot
  \equiv
  \bepsilon_a^\transpose \pback \varepsilon
  \ ,
\end{equation}
(or by
$\bepsilon_a \spinordot \equiv \bepsilon_a^\transpose \pback \varepsilon_{\rm alt}$
if the alternative spinor metric $\varepsilon_{\rm alt}$ is used).
The row spinor $\psi \spinordot$
corresponding to a column spinor $\psi = \psi^a \bepsilon_a$ is
\begin{equation}
\label{rowspinor}
  \psi \spinordot
  \equiv
  \psi^\transpose \pback \varepsilon
  =
  \psi^a \bepsilon_a \spinordot
  \ .
\end{equation}
Whereas a column spinor $\psi$ transforms under a rotor $R$
as~(\ref{rotorspinor}),
a row spinor $\psi \spinordot$ transforms as
\begin{equation}
\label{rotorrowspinor}
  R : \ 
  \psi \spinordot
  \equiv
  ( R \psi )^\transpose \pback \varepsilon
  \rightarrow
  \psi^\transpose \pback R^\transpose \pback \varepsilon
  =
  \psi^\transpose \pback \varepsilon \reverse{R}
  =
  \psi \cdot \reverse{R}
  \ .
\end{equation}


According to the usual rules of matrix multiplication,
it is possible to multiply a row spinor by a column spinor,
and a column spinor by a row spinor,
but not a row spinor by a row spinor,
nor a column spinor by a column spinor,
equations~(\ref{spinormultiplication}).
Rather than forbidding multiplication,
it is advantageous to consider the supergeometric algebra as consisting of
linear combinations of all species ---
scalars (= complex numbers), column spinors, row spinors, and multivectors ---
and setting the product of
column spinors with column spinors,
and
row spinors with row spinors,
to nothing.

The product of a row spinor $\psi \spinordot$
and a column spinor $\chi$ is a scalar (a complex number),
\begin{equation}
  \psi \cdot \chi
  =
  \psi^\transpose \pback \varepsilon \chi
  \ .
\end{equation}
Indeed the spinor metric $\varepsilon$ was constructed precisely so that
$\psi \cdot \chi$
transforms under a rotation as a scalar~(\ref{rotorscalarproduct}).
In components,
the scalar product of row and column spinors is
\begin{equation}
  \psi \cdot \chi
  =
  \psi^b \varepsilon_{ba} \chi^a
  \ .
\end{equation}
The scalar product is symmetric or antisymmetric
as the spinor metric $\varepsilon$ is symmetric or antisymmetric,
\begin{equation}
\label{symmetryspinorscalarproduct}
  \psi \cdot \chi
  =
  \pm
  \chi \cdot \psi
  \ ,
\end{equation}
the sign being as given in
Table~\ref{spinormetricsymmetrytab}.

The product of a column spinor $\chi$ and a row spinor $\psi \spinordot$
defines their outer product $\chi \psi \spinordot$.
The outer product transforms under a rotation
in the same way~(\ref{rotormultivector}) as a multivector,
\begin{equation}
  R : \ 
  \chi \psi \spinordot
  \equiv
  \chi \psi^\transpose \pback \varepsilon
  \rightarrow
  ( R \chi ) ( R \psi )^\transpose \pback \varepsilon
  =
  R ( \chi \psi \spinordot ) \reverse{R}
  \ .
\end{equation}
Multiplication of outer products satisfies the associative rule
\begin{equation}
\label{spinorassociative}
  ( \chi \psi \spinordot ) ( \varphi \xi \spinordot )
  =
  \chi ( \psi \cdot \varphi ) \xi \spinordot
  \ ,
\end{equation}
which since $\psi \cdot \varphi$ is a scalar
is proportional to the outer product $\chi \xi \spinordot$.
The associative rule~(\ref{spinorassociative}) makes it straightforward
to simplify long sequences of products of column and row spinors.
The trailing-dot notation for a row spinor makes the application
of the associative rule transparent.

\section{Algebra of outer products of spinors}
\label{algebraouterproduct-sec}

This section~\ref{algebraouterproduct-sec}
proves the Brauer-Weyl \cite{Brauer:1935} theorem,
that the algebra of outer products of spinors
is isomorphic to the geometric (Clifford) algebra of multivectors,
for the case of even spacetime dimensions $N$.
The case of odd dimensions $N$ is addressed in section~\ref{spinalgebraodd-sec}.
The claimed isomorphism
is established by an explicit representation
in terms of column and row vectors for spinors,
and matrices for multivectors in the geometric algebra.
Section~\ref{chiralrep-sec}
takes the spinor metric to be the standard spinor metric
$\varepsilon$, equation~(\ref{pspinormetricp}).
Then section~\ref{altpspinormetric-sec}
describes the modifications that must be made if the spinor metric
is taken to be the alternative spinor metric
$\varepsilon_{\rm alt}$, equation~(\ref{altpspinormetricp}).

Examples of the isomorphism between outer products of spinors
and the geometric algebra are given for the Pauli algebra ($N = 2$ or 3)
in Appendix~\ref{spinorgeom3D-app},
equations~(\ref{singlet}) and~(\ref{triplet}),
and for the Dirac algebra ($N = 4$)
in Appendix~\ref{spinorgeom31D-app},
equations~(\ref{singletR})--(\ref{quadrupletpseudodirac}).

Since $[N/2]$, the integer part of half the number of dimensions $N$,
appears so ubiquitously,
it is convenient to define
\begin{equation}
\label{nN}
  n \equiv [N/2]
  \ .
\end{equation}

\subsection{Chiral representation of the supergeometric algebra}
\label{chiralrep-sec}

The construction in this section~\ref{chiralrep-sec}
yields the chiral representation
of spinors and multivectors in even $N = 2n$ dimensions,
generated inductively starting from $N = 0$.
Given a representation of column and row basis spinors
$\bepsilon_A$ and $\bepsilon_A \spinordot$ in $N{-}2$ dimensions,
a representation of column and row basis spinors
$\bepsilon_{Aa}$ and $\bepsilon_{Aa} \spinordot$
(with one extra index $a$ = $\uparrow$ or $\downarrow$)
in $N$ dimensions are column and row matrices of length $2^n$,
\begin{subequations}
\label{bepsilonrep}
\begin{align}
\label{bepsilonrepup}
  \bepsilon_{A\uparrow}
  =
  \left(
  \begin{array}{c}
  \bepsilon_A \\ 0
  \end{array}
  \right)
  \ &, \quad
  \bepsilon_{A\uparrow} \spinordot
  =
  \left(
  \begin{array}{cc}
  0 & \bepsilon_A \spinordot
  \end{array}
  \right)
  \ ,
\\
\label{bepsilonrepdown}
  \bepsilon_{A\downarrow}
  =
  \left(
  \begin{array}{c}
  0 \\ \bepsilon_A
  \end{array}
  \right)
  \ &, \quad
  \bepsilon_{A\downarrow} \spinordot
  =
  \left(
  \begin{array}{cc}
  (-)^{n-1} \bepsilon_A \spinordot & 0
  \end{array}
  \right)
  \ ,
\end{align}
\end{subequations}
where $0$ represents respectively a zero column or row vector of
length $2^{n-1}$.
The induction starts at $n = 1$ where $A$ is empty
and $\bepsilon_A = \bepsilon_A \spinordot = 1$.
The trailing dot signifies the spinor metric tensor $\varepsilon$.
The construction~(\ref{bepsilonrep}) assumes that the spinor metric $\varepsilon$
is a product~(\ref{pspinormetricp}) of factors,
the last factor $\bgamma_n^+$ taking the form~(\ref{gammavecNrep}),
so that the relation between the spinor metric in $N$ and $N{-}2$ dimensions
is given by equation~(\ref{pspinormetricNN2}).

The outer products of the column basis spinors
$\bepsilon_{Aa}$
and row basis spinors
$\bepsilon_{Bb} \spinordot$
given by the inductive relations~(\ref{bepsilonrep}) are
$2^{n} \times 2^{n}$ matrices
\begin{subequations}
\label{bepsilontrixproto}
\begin{align}
  \bepsilon_{A\uparrow} \bepsilon_{B\uparrow} \spinordot
  &=
  \left(
  \begin{array}{cc}
  0 & \bepsilon_A \bepsilon_B \spinordot
  \\
  0 & 0
  \end{array}
  \right)
  \ ,
\\
  \bepsilon_{A\uparrow} \bepsilon_{B\downarrow} \spinordot
  &=
  \left(
  \begin{array}{cc}
  (-)^{n-1} \bepsilon_A \bepsilon_B \spinordot & 0
  \\
  0 & 0
  \end{array}
  \right)
  \ ,
\\
  \bepsilon_{A\downarrow} \bepsilon_{B\uparrow} \spinordot
  &=
  \left(
  \begin{array}{cc}
  0 & 0
  \\
  0 & \bepsilon_A \bepsilon_B \spinordot
  \end{array}
  \right)
  \ ,
\\
  \bepsilon_{A\downarrow} \bepsilon_{B\downarrow} \spinordot
  &=
  \left(
  \begin{array}{cc}
  0 & 0
  \\
  (-)^{n-1} \bepsilon_A \bepsilon_B \spinordot & 0
  \end{array}
  \right)
  \ ,
\end{align}
\end{subequations}
where the $0$'s
in equations~(\ref{bepsilontrixproto})
represent zero
$2^{n-1} \times 2^{n-1}$ matrices.
Again, the induction~(\ref{bepsilontrixproto})
starts at $n = 1$ where $A$ and $B$ are empty, and
$\bepsilon_A \bepsilon_B \spinordot = 1$.
The outer products~(\ref{bepsilontrixproto}) can be rewritten
\begin{subequations}
\label{bepsilontrix}
\begin{align}
\label{bepsilontrixpp}
  \bepsilon_{A\uparrow} \bepsilon_{B\uparrow} \spinordot
  &=
  \frac{1}{\sqrt{2}}
  \left(
  \begin{array}{cc}
  \bepsilon_A \bepsilon_B \spinordot & 0
  \\
  0 & \pm \bepsilon_A \bepsilon_B \spinordot
  \end{array}
  \right)
  \left(
  \begin{array}{cc}
  0 & \sqrt{2}
  \\
  0 & 0
  \end{array}
  \right)
  \ ,
\\
\label{bepsilontrixpm}
  \bepsilon_{A\uparrow} \bepsilon_{B\downarrow} \spinordot
  &=
  (-)^{n-1}
  \frac{1}{2}
  \left(
  \begin{array}{cc}
  \bepsilon_A \bepsilon_B \spinordot & 0
  \\
  0 & \pm \bepsilon_A \bepsilon_B \spinordot
  \end{array}
  \right)
  \left(
  \begin{array}{cc}
  2 & 0
  \\
  0 & 0
  \end{array}
  \right)
  \ ,
\\
\label{bepsilontrixmp}
  \bepsilon_{A\downarrow} \bepsilon_{B\uparrow} \spinordot
  &=
  \pm
  \frac{1}{2}
  \left(
  \begin{array}{cc}
  \bepsilon_A \bepsilon_B \spinordot & 0
  \\
  0 & \pm \bepsilon_A \bepsilon_B \spinordot
  \end{array}
  \right)
  \left(
  \begin{array}{cc}
  0 & 0
  \\
  0 & 2
  \end{array}
  \right)
  \ ,
\\
\label{bepsilontrixmm}
  \bepsilon_{A\downarrow} \bepsilon_{B\downarrow} \spinordot
  &=
  \pm (-)^{n-1}
  \frac{1}{\sqrt{2}}
  \left(
  \begin{array}{cc}
  \bepsilon_A \bepsilon_B \spinordot & 0
  \\
  0 & \pm \bepsilon_A \bepsilon_B \spinordot
  \end{array}
  \right)
  \left(
  \begin{array}{cc}
  0 & 0
  \\
  \sqrt{2} & 0
  \end{array}
  \right)
  \ ,
\end{align}
\end{subequations}
where the upper/lower sign is for even/odd
$\bepsilon_A \bepsilon_B \spinordot$
(that is, the total charge
of $\bepsilon_A \bepsilon_B \spinordot$ is even/odd;
multivectors of even and odd grade are respectively even and odd).
The first matrix on the right hand sides of equations~(\ref{bepsilontrix})
is the matrix representation of the multivector $\bepsilon_A \bepsilon_B \spinordot$
in $N$ dimensions in terms of its representation in $N{-}2$ dimensions,
\begin{equation}
\label{bepsilontrixNN2}
  \bepsilon_A \bepsilon_B \spinordot
  =
  (-)^{n-1}
  \bepsilon_{A\uparrow} \bepsilon_{B\downarrow} \spinordot
  \pm
  \bepsilon_{A\downarrow} \bepsilon_{B\uparrow} \spinordot
  =
  \left(
  \begin{array}{cc}
  \bepsilon_A \bepsilon_B \spinordot & 0
  \\
  0 & \pm \bepsilon_A \bepsilon_B \spinordot
  \end{array}
  \right)
  \ .
\end{equation}
The rightmost factors in equations~(\ref{bepsilontrix})
constitute the matrix representations of the chiral multivectors
$\bgamma_n$, $\bgamma_n \bgamma_{\bar n}$, $\bgamma_{\bar n} \bgamma_n$, and $\bgamma_{\bar n}$
in $N = 2n$ dimensions,
\begin{subequations}
\label{gammapmrep}
\begin{align}
  \bgamma_n
  &=
  \left(
  \begin{array}{cc}
  0 & \sqrt{2}
  \\
  0 & 0
  \end{array}
  \right)
  \ , \quad
  \bgamma_n \bgamma_{\bar n}
  =
  \left(
  \begin{array}{cc}
  2 & 0
  \\
  0 & 0
  \end{array}
  \right)
  \ ,
\\
  \bgamma_{\bar n}
  &=
  \left(
  \begin{array}{cc}
  0 & 0
  \\
  \sqrt{2} &0
  \end{array}
  \right)
  \ , \quad
  \bgamma_{\bar n} \bgamma_n
  =
  \left(
  \begin{array}{cc}
  0 & 0
  \\
  0 & 2
  \end{array}
  \right)
  \ ,
\end{align}
\end{subequations}
which have the correct normalization and commutation rules with
respect to each other.
The signs in equations~(\ref{bepsilontrix})
are arranged so that the correct commutation rules of
the geometric algebra are respected:
$\bgamma_n$ and $\bgamma_{\bar n}$, which are odd,
commute/anticommute with
$\bepsilon_A \bepsilon_B \spinordot$
according as the latter is even/odd;
and
$\bgamma_n \bgamma_{\bar n}$ and $\bgamma_{\bar n} \bgamma_n$, which are even,
always commute with
$\bepsilon_A \bepsilon_B \spinordot$.
In terms of scalar and wedge products,
the multivectors
$\bgamma_n \bgamma_{\bar n}$ and $\bgamma_{\bar n} \bgamma_n$
in equations~(\ref{gammapmrep}) are
\begin{subequations}
\begin{alignat}{3}
\label{sgmapm}
  &
  \bgamma_n \bgamma_{\bar n}
  =
  \bgamma_n \cdot \bgamma_{\bar n}
  +
  \bgamma_n \wedgie \bgamma_{\bar n}
  \ , \quad
  &&
  \bgamma_{\bar n} \bgamma_n
  =
  \bgamma_n \cdot \bgamma_{\bar n}
  -
  \bgamma_n \wedgie \bgamma_{\bar n}
  \ ,
\\
  &
  \bgamma_n \cdot \bgamma_{\bar n}
  =
  \left(
  \begin{array}{cc}
  1 & 0
  \\
  0 & 1
  \end{array}
  \right)
  \ , \quad
  &&
  \bgamma_n \wedgie \bgamma_{\bar n}
  =
  \left(
  \begin{array}{cc}
  1 & 0
  \\
  0 & -1
  \end{array}
  \right)
  \ .
\end{alignat}
\end{subequations}
Note that chiral and orthonormal bivectors are related by
\begin{equation}
  \bgamma_n \wedgie \bgamma_{\bar n} = - \im \, \bgamma_n^+ \wedgie \bgamma_n^-
  \ ,
\end{equation}
so that $( \bgamma_n \wedgie \bgamma_{\bar n} )^2 = 1$.
The orthonormal basis vectors $\bgamma_n^+$ and $\bgamma_n^-$
at the $n$'th step are
\begin{equation}
\label{gammavecNrep}
  \bgamma_n^+
  =
  \left(
  \begin{array}{cc}
  0 & 1
  \\
  1 & 0
  \end{array}
  \right)
  \ , \quad
  \bgamma_n^-
  =
  \left(
  \begin{array}{cc}
  0 & -\im
  \\
  \im & 0
  \end{array}
  \right)
  \ ,
\end{equation}
which are traceless, unitary, and Hermitian.
The orthonormal basis bivector
$\bgamma_n^+ \wedgie \bgamma_n^-$
is
\begin{equation}
\label{gammabivecNrep}
  \bgamma_n^+ \wedgie \bgamma_n^-
  =
  \left(
  \begin{array}{cc}
  \im & 0
  \\
  0 & -\im
  \end{array}
  \right)
  \ ,
\end{equation}
which is traceless, unitary, and skew-Hermitian.
An iterative expression for the spinor metric
$\varepsilon_{N}$
follows from its expression~(\ref{pspinormetricp})
as a product of basis vectors, and is the antidiagonal matrix
\begin{align}
\label{pspinormetricNN2}
  \varepsilon_{N}
  =
  \varepsilon_{N-2} \,
  \bgamma_n^+
  &=
  \left(
  \begin{array}{cc}
  \varepsilon_{N-2} & 0
  \\
  0 & (-)^{n-1} \varepsilon_{N-2}
  \end{array}
  \right)
  \left(
  \begin{array}{cc}
  0 & 1
  \\
  1 & 0
  \end{array}
  \right)
\nonumber
\\
  &=
  \left(
  \begin{array}{cc}
  0 & \varepsilon_{N-2}
  \\
  (-)^{n-1} \varepsilon_{N-2} & 0
  \end{array}
  \right)
  \ .
\end{align}
The left factor in the rightmost expression in the top row
of equations~(\ref{pspinormetricNN2})
is the matrix representation of $\varepsilon_{N-2}$ in $N$ dimensions
in terms of its representation in $N{-}2$ dimensions,
in accordance with equation~(\ref{bepsilontrixNN2}).
The factor of $(-)^{n-1}$
comes from the fact that the spinor metric $\varepsilon_{N-2}$
is a product of $n{-}1$ basis vectors,
equation~(\ref{pspinormetricp}),
so is even/odd (total charge even/odd) as $n{-}1$ is even or odd.
Equation~(\ref{pspinormetricNN2}),
which was assumed in the initial step~(\ref{bepsilonrep}) of the construction
of the chiral representation of the supergeometric algebra,
proves the consistency of the construction.

The matrix representation of
the column and row basis spinors~(\ref{bepsilonrep})
and of their outer products~(\ref{bepsilontrix})
is entirely real (with respect to $\im$).
The expansion of the $2^N$ outer products
$\bepsilon_a \bepsilon_b \spinordot$
of spinors in terms of
the $2^N$ basis multivectors $\bgamma_A$ of the spacetime algebra,
and vice versa,
define the matrix of coefficients
$\gamma^A_{ab}$
and its inverse
$\gamma_A^{ab}$,
\begin{equation}
  \bepsilon_a \bepsilon_b \spinordot
  =
  \gamma^A_{ab}
  \bgamma_A
  \ , \quad
  \bgamma_A
  =
  \gamma_A^{ab}
  \bepsilon_a \bepsilon_b \spinordot
  \ .
\end{equation}
The coefficients
$\gamma^A_{ab}$
and
$\gamma_A^{ab}$
in the chiral representation
are
\begin{equation}
\label{gammacoeff}
  \gamma^A_{ab}
  =
  {1 \over 2^{n}} \,
  \bepsilon_b \cdot \bgamma^A \bepsilon_a
  \ , \quad
  \gamma_A^{ab}
  =
  \sign ( \varepsilon^2 ) \,
  \bepsilon^a \cdot \bgamma_A \bepsilon^b
  \ ,
\end{equation}
where
$\sign ( \varepsilon^2 )$ is the symmetry of the spinor metric,
Table~\ref{spinormetricsymmetrytab}.

The trace of the outer product of a pair of basis spinors
gives their scalar product,
\begin{equation}
  \Tr \,
  \bepsilon_a \, \bepsilon_b \spinordot
  =
  \bepsilon_b \cdot \bepsilon_a
  =
  \varepsilon_{ba}
  \ .
\end{equation}
More generally, the trace of an outer product of spinors equals
their scalar product,
\begin{equation}
  \Tr \,
  \chi \psi \spinordot
  =
  \psi \cdot \chi
  \ .
\end{equation}

The above construction
establishes the claimed isomorphism between
outer products of spinors and the geometric algebra in even dimensions $N$,
\begin{equation}
  \mbox{outer products of spinors}
  \,\cong\,
  \mbox{geometric algebra}
  \quad
  \mbox{($N$ even)}
  \ .
\end{equation}
Both spaces are complex $2^N$-dimensional vector spaces.
Their representation as $2^{n} \times 2^{n}$ dimensional matrices
is minimal:
there is no representation of the geometric algebra
with matrices of smaller dimension.

\subsection{Chiral representation of the supergeometric algebra using the alternative spinor metric}
\label{altpspinormetric-sec}

The chiral representation of the supergeometric algebra with
the alternative spinor metric~(\ref{altpspinormetricp})
is the same as the construction in section~\ref{chiralrep-sec},
but with the replacement
\begin{equation}
  (-)^{n-1}
  \rightarrow
  (-)^{n}
\end{equation}
in equations~(\ref{bepsilonrep}) to~(\ref{bepsilontrixNN2}).
Analogously to equation~(\ref{pspinormetricNN2}),
an iterative equation for the alternative spinor
metric follows from its expression~(\ref{altpspinormetricp})
as a product of basis vectors,
and is the antidiagonal matrix
\begin{align}
\label{altpspinormetricNN2}
  \varepsilon^{\rm alt}_{N}
  =
  \varepsilon^{\rm alt}_{N-2} \,
  \im \bgamma_n^-
  &=
  \left(
  \begin{array}{cc}
  \varepsilon^{\rm alt}_{N-2} & 0
  \\
  0 & (-)^{n-1} \varepsilon^{\rm alt}_{N-2}
  \end{array}
  \right)
  \left(
  \begin{array}{cc}
  0 & 1
  \\
  -1 & 0
  \end{array}
  \right)
\nonumber
\\
  &=
  \left(
  \begin{array}{cc}
  0 & \varepsilon^{\rm alt}_{N-2}
  \\
  (-)^{n} \varepsilon^{\rm alt}_{N-2} & 0
  \end{array}
  \right)
  \ .
\end{align}

\subsection{Properties of orthonormal basis multivectors in the chiral representation}
\label{basisproperties-sec}

In the chiral representation constructed
in sections~\ref{chiralrep-sec} or~\ref{altpspinormetric-sec},
all orthonormal basis vectors $\bgamma_a$,
and all orthonormal basis $p$-vectors
$\bgamma_{a_1 ... a_p} \equiv \bgamma_{a_1} \wedgie ... \wedgie \bgamma_{a_p}$,
are traceless (except for the unit basis element 1),
unitary,
and either Hermitian (if $[p/2]$ is even, i.e.\ $p = 0, 1, 4, 5, ...$)
or skew-Hermitian (if $[p/2]$ is odd, i.e.\ $p = 2, 3, 6, 7, ...$)
$2^n \times 2^n$ matrices.
All matrices have determinant 1,
except that for $N = 2$ the vectors (grade $p = 1$) have determinant $-1$.
The unit element is represented by the unit matrix.
Most of these assertions can be proved
by induction using the expression~(\ref{bepsilontrixNN2}),
which gives the representation of a multivector in $N$ dimensions
in terms of its representation in $N{-}2$ dimensions.

\section{Supergeometric algebra in odd dimensions}
\label{spinalgebraodd-sec}

The natural complex structure of spinors means that supergeometric algebras
are naturally even-dimensional.
Section~\ref{algebraouterproduct-sec}
proved the Brauer-Weyl theorem for even dimensions.
What about odd dimensions $N$?

There are essentially two options.
One option, considered in section~\ref{oddN-sec},
is to project the odd $N$-dimensional geometric algebra
into one lower dimension.
The other option, considered in section~\ref{oddN2-sec},
is to embed the odd $N$-dimensional geometric algebra in one higher dimension.

\subsection{Supergeometric algebra in odd dimensions, version 1}
\label{oddN-sec}

The first approach,
projecting the odd-dimensional algebra into one lower dimension,
requires identifying the chiral operator $\varkappa_N$ with 1,
equation~(\ref{kappaNodd}).
This is the standard convention for geometric algebras in odd dimensions,
including the Pauli algebra, $N = 3$.


The pseudoscalar $I_N$ of the geometric algebra,
in either even or odd dimensions $N$,
is the product of all $N$ orthonormal vectors
(the orthonormal vectors $\bgamma_k^\pm$ and $\bgamma_N$
here are being taken to be spacelike,
as in section~\ref{algebraouterproduct-sec};
if some of the vectors are timelike,
then the usual physics convention for the pseudoscalar
would have extra factors of $\im$, one for each timelike dimension,
but for simplicity that adjustment is not made here),
\begin{equation}
\label{pseudoscalarN}
  I_N
  \equiv
  \bgamma_1^+ \wedgie \bgamma_1^- \wedgie ... \wedgie \bgamma_n^+ \wedgie \bgamma_n^-
  \left\{
  \wedgie \bgamma_N
  \mbox{ if $N$ is odd}
  \right\}
  =
  \im^{n}
  \varkappa_N
\end{equation}
(with $n \equiv [N/2]$ as before).
The chiral operator $\varkappa_N$
is the pseudoscalar normalized by a phase so that its square is the unit matrix,
\begin{equation}
\label{kappaN}
  \varkappa_N
  \equiv
  \bgamma_1 \wedgie \bgamma_{\bar 1} \wedgie
  ...
  \wedgie \bgamma_n \wedgie \bgamma_{\bar n}
  \left\{
  \wedgie \bgamma_N
  \mbox{ if $N$ is odd}
  \right\}
  \ .
\end{equation}
In the chiral representation~(\ref{bepsilontrix}),
the representation of the chiral operator $\varkappa_N$ in $N$ even dimensions
in terms of its representation $\varkappa_{N-2}$ in $N{-}2$ dimensions is
the diagonal matrix
\begin{equation}
\label{kappaNN2}
  \varkappa_N
  =
  \left(
  \begin{array}{cc}
  \varkappa_{N-2} & 0 \\
  0 & - \varkappa_{N-2}
  \end{array}
  \right)
  \quad
  \mbox{($N$ even)}
  \ .
\end{equation}
The chiral operator is diagonal in the chiral representation by construction.
The square of the pseudoscalar is $I_N^2 = (-)^n$,
so the square of the chiral operator is the unit matrix~$1$,
\begin{equation}
\label{kappaN2}
  \varkappa_N^2 = 1
  \ .
\end{equation}
Like the pseudoscalar $I_N$,
the chiral operator $\varkappa_N$ is invariant under rotations.

For even $N$,
the chiral operator $\varkappa_N$ is defined through equation~(\ref{kappaN})
as a prescribed member of both algebras,
the algebra of spinor outer products and the geometric algebra.
But for odd $N$,
since the definition~(\ref{kappaN}) involves $\bgamma_N$
which has (as yet) no expression in the algebra of outer products of spinors,
there is the possibility that $\varkappa_N$ could be a distinct element
not belonging to the algebra of spinor outer products.
The element $\varkappa_N$ is a rotationally invariant scalar that squares to 1,
and that, for odd $N$, commutes with all basis vectors
$\bgamma_k^\pm$, $k = 1 , ... , n$ other than $\bgamma_N$.
The other element of the odd-$N$ algebra of spinor outer products that
possesses those properties is (up to a possible sign) the unit element.
Thus if the chiral operator $\varkappa_N$ is identified with $1$,
\begin{equation}
\label{kappaNodd}
  \varkappa_N = 1
  \quad
  \mbox{($N$ odd)}
  \ ,
\end{equation}
then there is an isomorphism between the algebra of outer products
of spinors in $N{-}1$ dimensions
and the geometric algebra in $N$ dimensions
modulo the chiral operator $\varkappa_N$,
\begin{equation}
  \mbox{outer products of spinors}
  \,\cong\,
  \mbox{geometric algebra (${\rm mod}$ $\varkappa_N$)}
  \quad
  \mbox{($N$ odd)}
  \ .
\end{equation}
Given the identification~(\ref{kappaNodd}) of the chiral operator with $1$,
it then follows from the definition equation~(\ref{kappaN}) of $\varkappa_N$
that the final element $\bgamma_N$ of the geometric algebra is
\begin{equation}
\label{gammaNodd}
  \bgamma_N
  =
  \varkappa_{N-1}
  =
  \bgamma_1 \wedgie \bgamma_{\bar 1} \wedgie
  ...
  \wedgie \bgamma_n \wedgie \bgamma_{\bar n}
  \quad
  \mbox{($N$ odd)}
  \ .
\end{equation}
In the case $N = 3$, this gives
\begin{equation}
\label{gamma3odd}
  \bgamma_3
  =
  \bgamma_1 \wedgie \bgamma_{\bar 1}
  =
  \left(
  \begin{array}{cc}
  1 & 0
  \\
  0 & -1
  \end{array}
  \right)
  \ ,
\end{equation}
in agreement with the usual expression for the third Pauli matrix.
With the identification~(\ref{kappaNodd}),
the pseudoscalar $I_N$ itself is, equation~(\ref{pseudoscalarN}),
\begin{equation}
\label{IiNodd}
  I_N
  =
  \im^n
  \quad
  \mbox{($N$ odd)}
  \ .
\end{equation}
The final basis vector $\bgamma_N$, equation~(\ref{gammaNodd}),
of the odd algebra
has the same properties as the other orthonormal basis vectors
$\bgamma_k^\pm$, $k = 1 , ... , n$:
its square is 1,
it anticommutes with the other orthonormal basis vectors,
it is represented by a traceless, unitary, Hermitian matrix,
and its reverse is (by definition) itself,
$\reverse{\bgamma}_N = \bgamma_N$.
And,
like the other orthonormal basis vectors $\bgamma_k^+$ of odd index,
the representation of $\bgamma_N$ is real.

For odd $N$, the chiral operator $\varkappa_N$
defined by equation~(\ref{kappaN})
is (before $\varkappa_N$ is identified with 1)
an odd element of the geometric algebra.
Thus for odd $N$, the odd part of the geometric algebra
is isomorphic to $\varkappa_N$ times the even geometric algebra.
Only the odd geometric algebra is affected
by the identification~(\ref{kappaNodd}) of the chiral operator with unity;
the even geometric algebra is unaffected.
The square of the chiral operator is always 1, equation~(\ref{kappaN2}),
so the product of two odd multivectors yields the correct even multivector
regardless of the identification~(\ref{kappaNodd}).



In summary,
the algebra of spinor outer products in $2[N/2]$ dimensions
is isomorphic to the geometric algebra for both even and odd $N$,
modulo $\varkappa_N$ in the case of odd $N$.
The algebra is a complex (with respect to $\im$)
vector space of dimension $2^n$,
represented in the chiral construction~(\ref{bepsilontrix})
by $2^n \times 2^n$ matrices.
For example,
the $N = 2$ geometric algebra is the algebra generated by
$1, \bgamma_1^+ , \bgamma_1^- , \bgamma_1^+ \wedgie \bgamma_1^-$,
while the $N = 3$ geometric algebra (the Pauli algebra)
is the algebra generated by
$1, \bgamma_1^+ , \bgamma_1^- , \bgamma_3$,
the pseudoscalar $I_3$ being identified with the imaginary $\im$.

\subsection{Extra symmetry of the supergeometric algebra in odd dimensions}
\label{oddsymmetry-sec}

Given that,
if $\varkappa_N$ is identified with 1,
the geometric algebra for odd $N$ is isomorphic to
the geometric algebra for even $N{-}1$,
what is the difference between the two algebras?
Since the algebras are isomorphic, there is of course no difference.
However, bivectors are special in that they are the only
generators that generate transformations that preserve grade,
and therefore correspond to what one usually thinks of as rotations.
If one restricts to rotations generated by bivectors,
then the odd algebra has a higher degree of symmetry.
The equivalence~(\ref{gammaNodd}) means that
the pseudoscalar $\varkappa_{N-1}$ in the even algebra
is promoted to a vector $\bgamma_N$ in the odd algebra,
and
pseudovectors $\bgamma_k^\pm \varkappa_{N-1}$ in the even algebra
become bivectors $\bgamma_k^\pm \wedgie \bgamma_N$ in the odd algebra.
Thus the odd geometric algebra has $N{-}1$ more rotations than the even algebra.

The Pauli algebra in $N = 3$ dimensions
offers a familiar example.
In both 2 and 3 dimensions there are just 2 basis spinors,
$\bepsilon_\uparrow$ and $\bepsilon_\downarrow$,
which one commonly conceptualizes as being up and down along a ``3-axis''.
But whereas in 2 dimensions there is just one rotation,
generated by the bivector $\bgamma_1^+ \wedgie \bgamma_1^-$
(rotation about the ``3-axis''),
in 3 dimensions there are 2 more rotations,
generated by the bivectors
$\bgamma_1^+ \wedgie \bgamma_3$ and $\bgamma_1^- \wedgie \bgamma_3$
(rotations about the ``2-axis'' and ``1-axis'').

\subsection{Parity and time reversal}
\label{paritytime-sec}

Parity reversal $P$ is the operation of reflecting an odd number
of spatial axes.
Time reversal $T$ is the operation of reflecting an odd number
of time axes.
The combination $P T$ reflects an odd number of spatial axes
and an odd number of time axes.
Parity and tine reversal are discrete transformations that cannot be
achieved by any continuous rotation.
A reflection of an even number of spatial axes can
be accomplished by a continuous rotation in spatial dimensions,
while a reflection of an even number of time axes can
be accomplished by a continuous rotation in time dimensions.

\paritytimetab

Consider a transformation $X$
that transforms spinors $\psi$ and multivectors $\ba$ according to
\begin{equation}
\label{Xtransformation}
  X : \quad
  \psi \rightarrow X \psi
  \ , \quad
  \ba \rightarrow X \ba X^{-1}
  \ .
\end{equation}
The form~(\ref{Xtransformation}) of the $X$ transformation,
which is similar to the form of transformations~(\ref{rotormultivector})
and~(\ref{rotorspinor}) under rotations,
ensures consistency between transformations of spinors and multivectors.
If $X$ is set equal to an orthonormal dimension $\bgamma_a$,
then the transformation~(\ref{Xtransformation}) flips all orthonormal axes
{\em except\/} $\bgamma_a$.
This assertion remains true
in the situation considered in section~\ref{oddN-sec},
where the number of $N$ of spacetime dimensions is odd,
and the final vector $\bgamma_N$ is identified with the $(N{-}1)$-dimensional
chiral operator, equation~(\ref{gammaNodd}).

Table~\ref{paritytimetab} summarizes
how setting the operator $X$ in the transformation~(\ref{Xtransformation})
either to a spacelike orthonormal basis vector $\bgamma_a$,
or to a timelike orthonormal basis vector $\bgamma_m$,
yields a parity transformation $P$, a time reversal transformation $T$,
a product $PT$ of both, or neither.
For example,
if $X$ is a spacelike orthonormal basis vector $\bgamma_a$,
then the transformation~(\ref{Xtransformation})
flips $K{-}1$ of the $K$ spacelike axes, and $M$ of the $M$ timelike axes,
so flips $P$ if $K$ is even, and $T$ if $M$ is odd;
and similarly
if $X$ is a timelike orthonormal basis vector $\bgamma_m$,
then the transformation~(\ref{Xtransformation})
flips $K$ of the $K$ spacelike axes, and $M{-}1$ of the $M$ timelike axes,
so flips $P$ if $K$ is odd, and $T$ if $M$ is even.
Setting $X$ equal to products of spacelike/timelike basis vectors
yields corresponding products of $P$, $T$, $PT$, or $-$.
As Table~\ref{paritytimetab} shows,
individual $P$ and $T$ operators are available only
if the total number $N = K{+}M$ of spacetime dimensions is even.
If on the other hand
the total number $N = K{+}M$ of spacetime dimensions is odd,
then only a combination $PT$ operator is available
(provided that there is at least one spatial dimension
and at least one time dimension).

A solution to the problem of constructing an odd-$N$ supergeometric algebra
that incorporates individual parity and time reversal operators
is presented in the next section~\ref{oddN2-sec}.

\subsection{Supergeometric algebra in odd dimensions, version 2}
\label{oddN2-sec}

The previous section~\ref{paritytime-sec}
brought up the fact that the
geometric algebra in odd $N$ dimensions does not contain
a parity operator $P$ or time reversal operator $T$.

The problem is not that the operations of parity or time reversal do not exist,
but rather, how to construct such operators as elements of the supergeometric algebra.

A solution is to embed the odd $N$-dimensional algebra
in the even $(N{+}1)$-dimensional algebra,
and to treat
either the final (odd) dimension $\bgamma_N$
or the extra (even) orthonormal dimension $\bgamma_{N+1}$
as a scalar dimension.
The dimension is scalar in the sense that,
although it transforms under rotations in the enlarged group $\Spin(N{+}1)$,
it is fixed under rotations in the $\Spin(N)$ subgroup of the enlarged group.
Setting $X$ equal to the scalar dimension,
\begin{equation}
\label{sgmaarbitrarydimensionparitydef}
  X
  =
  \bgamma_N
  ~\mbox{or}~
  \bgamma_{N+1}
  \ ,
\end{equation}
yields an operator that flips the $N$ other orthonormal vectors.
Since $N$ is odd, $X$ flips an odd number of orthonormal vectors.
$X$ is a parity operator $P$ if the number $K$ of spatial dimensions is odd,
or a time reversal operator $T$ if $K$ is even.

As usual, there is a spin bit (the $[(N{+}1)/2]$'th bit) associated with
the pair $\bgamma_N$ and $\bgamma_{N+1}$ of axes.
Normally a rotation in the $\bgamma_N \wedgie \bgamma_{N+1}$ plane
would rotate spinors by a phase $\ee^{\mp \im \theta / 2}$
with sign $\mp$ depending on whether the spin bit is
up $\uparrow$ or down $\downarrow$.
But since one of $\bgamma_N$ or $\bgamma_{N+1}$ is a scalar,
there is no such rotation.
Notwithstanding the absence of a rotation by a phase,
the spin bit is still there, part of the bitcode index
$a = a_1 ... a_{[(N+1)/2]}$ of a basis spinor $\bepsilon_a$.

\section{Chiral subalgebras}
\label{chiral-sec}

Chirality,
the eigenvalue of the chiral operator $\varkappa_N$,
counts whether the number of up-bits of a spinor is even or odd,
with right-handed defined as the all-bit-up spinor.
In even spacetime dimensions $N$,
rotations preserve chirality.
\cite{Cartan:1938} refers to spinors of definite (right- or left-handed)
chirality as semi-spinors.
In 3+1 spacetime dimensions,
Dirac spinors of definite chirality are called Weyl spinors.


In odd $N$ dimensions,
if the path proposed in section~\ref{oddN-sec} is followed,
where the algebra is projected into one lower dimension,
which requires identifying the chiral operator $\varkappa_N$ with unity,
equation~(\ref{kappaNodd}),
then rotations mix right-and left-handed spinors,
and chirality is not a rotationally invariant property of spinors.

If on the other hand $N$ is odd and
the path proposed in section~\ref{oddN2-sec} is followed,
where the algebra is embedded in one higher dimension,
then a basis spinor $\bepsilon_a$ has $[(N{+}1)/2]$ bits,
and chirality is that $\varkappa_{N+1}$ of the algebra in one higher dimension.
In the rest of this section~\ref{chiral-sec},
replace $N$ by $N{+}1$ if $N$ is odd
and section~\ref{oddN2-sec} is followed.

Right- and left-handed chiral multivectors
are eigenvalues of the chiral operator $\varkappa_N$
(or $\varkappa_{N+1}$ if $N$ is odd
and section~\ref{oddN2-sec} is followed),
with eigenvalues $\pm 1$,
\begin{equation}
  \varkappa_N
  \ba_{\overset{\scriptstyle\Rchiral}{\scriptstyle\Lchiral}}
  =
  \pm
  \ba_{\overset{\scriptstyle\Rchiral}{\scriptstyle\Lchiral}}
  \ .
\end{equation}
Right- and left-handed chirality projection operators
$P_{\overset{\scriptstyle\Rchiral}{\scriptstyle\Lchiral}}$
may be defined by
\begin{equation}
  P_{\overset{\scriptstyle\Rchiral}{\scriptstyle\Lchiral}}
  \equiv
  \tfrac{1}{2} ( 1 \pm \varkappa_N )
  =
  \tfrac{1}{2} ( 1 \pm \im^{-n} I_N )
  \ ,
\end{equation}
which are projection operators because their squares are one,
$(P_\Rchiral )^2 = (P_\Lchiral )^2 = 1$,
and their product is zero,
$P_\Rchiral P_\Lchiral = 0$.
A multivector $\ba$ splits into right- and left-handed chiral parts,
\begin{equation}
  \ba
  =
  \ba_\Rchiral + \ba_\Lchiral
  , \quad
  \ba_{\overset{\scriptstyle\Rchiral}{\scriptstyle\Lchiral}}
  \equiv
  P_{\overset{\scriptstyle\Rchiral}{\scriptstyle\Lchiral}} \ba
  \ .
\end{equation}
Since the chiral operator $\varkappa_N$
is proportional to the pseudoscalar $I_N$,
a purely right- or left-handed multivector is necessarily
a linear combination of a multivector $\ba$ and its Hodge dual $I_N \ba$.

An outer product of a right-handed column spinor with any row spinor
(right- or left-handed)
is a right-handed multivector.
An outer product of a left-handed column spinor with any row spinor
is a left-handed multivector.

\section{Conjugation}
\label{conjugation-sec}

Supergeometric algebras have a natural complex structure.
Complex conjugation is a discrete operation that
transforms a spinor into its antispinor partner.
However, complex conjugation of a spinor
is by itself not a Lorentz-invariant operation.
The conjugation operator $\Cp$ is introduced
to provide a Lorentz-invariant version of complex conjugation.

\subsection{Conjugation operator}
\label{conjugationop-sec}

In the chiral representation,
chiral basis spinors $\bepsilon_a$ are real,
and the complex conjugate of a spinor $\psi \equiv \psi^a \bepsilon_a$ is
\begin{equation}
  \psi^\ast \equiv \psi^{a \ast} \bepsilon_a
  \ .
\end{equation}
The complex conjugate spinor $\psi^\ast$ transforms under a rotor $R$
not like the spinor $\psi$, but rather as
\begin{equation}
  R : \ 
  \psi^\ast \rightarrow R^\ast \psi^\ast
  \ .
\end{equation}
The rotationally-invariant conjugation operator $\Cp$ is defined such that
commutation with it converts rotors $R$
to their complex conjugates,
\begin{equation}
\label{CpRN}
  \Cp R^\ast = R \Cp
  \ .
\end{equation}
Note that since a rotor $R$ is a real linear combination of even orthonormal
basis multivectors,
the complex conjugate $R^\ast$ of a rotor $R$ is a rotor.
The conjugate $\conj{\psi}$ of a spinor is defined by
\begin{equation}
\label{conjspinor}
  \conj{\psi}
  \equiv
  \Cp \psi^\ast
  \ .
\end{equation}
Thanks to the condition~(\ref{CpRN}) on the conjugation operator $\Cp$,
the conjugate spinor $\conj{\psi}$ transforms under a rotor $R$
in the same way as the spinor $\psi$,
\begin{equation}
  R : \ 
  \conj{\psi}
  \equiv
  \Cp \psi^\ast
  \rightarrow
  \Cp ( R \psi )^\ast
  =
  \Cp R^\ast \psi^\ast
  =
  R \Cp \psi^\ast
  =
  R \conj{\psi}
  \ .
\end{equation}

If all dimensions are spacelike,
then rotors are unitary, $R^{-1} = R^\dagger$,
in which case
the condition~(\ref{CpRN}) on the conjugation operator $\Cp$
is the same as the condition~(\ref{pspinormetrictensorN})
on the spinor metric $\varepsilon$,
and the conjugation operator is just equal to the spinor metric,
$\Cp = \varepsilon$
(or to the alternative spinor metric,
$\Cp_{\rm alt} = \varepsilon_{\rm alt}$).
But if there are $M$ timelike dimensions,
then each timelike dimension can be treated as
$\im$ times a spacelike dimension,
introducing an extra minus sign in the condition~(\ref{CpRN})
compared to the spacelike case, for each timelike factor.
The result is that the conjugation operator
$\Cp$ equals,
modulo a normalization factor,
the product of the spinor metric tensor $\varepsilon$
(or the alternative spinor metric tensor $\varepsilon_{\rm alt}$)
with the product of all timelike orthonormal basis vectors,
\begin{equation}
\label{arbdimensionCp}
  \Cp
  =
  \varepsilon \, \bGamma^\transpose
  \ , \quad
  \bGamma
  \equiv
  c
  \prod_{m}
  \bgamma_m
  (\mbox{timelike})
  \ ,
\end{equation}
where the phase factor $c$ is chosen
such that the square of $\bGamma$ is the unit matrix,
\begin{equation}
\label{CpGamma2}
  \bGamma^2 = 1
  \ .
\end{equation}
The conjugation operator $\Cp$ is unitary,
as is $\bGamma$.
The condition~(\ref{CpGamma2}) implies that
the eigenvalues of $\bGamma$ are $\pm 1$,
which ensures that $\conj{\psi} \cdot \psi$ is real,
equation~(\ref{spinorscalar}).
If there is at least one time dimension
(and, if $N$ is odd and section~\ref{oddN-sec} is followed,
at least one spatial dimension),
then the trace of $\bGamma$ is zero,
implying that there are equal numbers of $+1$ and $-1$ eigenvalues.
If there is just one time dimension $\bgamma_0$,
then $\bGamma$ is
(the minus sign is optional, but conventional in the Dirac algebra)
\begin{equation}
\label{arbdimensiongam0}
  \bGamma
  =
  - \im \bgamma_0
  \ .
\end{equation}
Notwithstanding equation~(\ref{arbdimensionCp}),
the conjugation operator $\Cp$ is defined by equation~(\ref{CpRN}),
and does not transform as an element of the geometric algebra,
but rather is a spinor tensor constructed such that conjugation,
equation~(\ref{conjspinor}),
is a Lorentz-invariant operation.

\cpsquaretab

The double conjugate of a spinor is
\begin{equation}
\label{conjconjbasisspinorKM}
  \conj{\conj{\psi}}
  =
  \Cp ( \Cp \psi^\ast )^\ast
  =
  \Cp \Cp^\ast \psi
  =
  \pm \psi
  \ ,
\end{equation}
where the sign is $+$ or $-$ depending on whether the conjugation operator
is symmetric, $\Cp = \Cp^\transpose$, or antisymmetric
(the symmetry condition $\Cp = \Cp^\transpose$ is
equivalent to $\Cp \Cp^\ast = 1$ in view of the unitarity of $\Cp$).
Table~\ref{cpsquaretab}
shows the symmetry of the conjugation operator $\Cp$
for the standard and alternative spinor metrics.
Table~\ref{cpsquaretab}
is essentially identical to the earlier
Table~\ref{spinormetricsymmetrytab},
except that the number $N = K{+}M$ of spacetime dimensions
in Table~\ref{spinormetricsymmetrytab}
is changed to the difference $K{-}M$
of numbers $K$ and $M$ of space and time dimensions
in Table~\ref{cpsquaretab}.
For Dirac spinors in 3+1 dimensions, for which $K{-}M = 3{-}1 = 2$,
the conventional choice is
the standard spinor metric~(\ref{pspinormetricp}),
which ensures that the conjugation operator is symmetric,
hence that the double conjugate of a Dirac spinor is itself,
$\conj{\conj{\psi}} = \psi$.

The scalar product of a conjugate spinor $\conj{\psi}$
with a spinor $\chi$ is
\begin{equation}
  \conj{\psi} \cdot \chi
  =
  \psi^\dagger \Cp^\transpose \pback \varepsilon \chi
  =
  \psi^\dagger \bGamma \chi
  \ ,
\end{equation}
which is a complex (with respect to $\im$) number.
The scalar product of a spinor $\psi$ with its conjugate is
\begin{equation}
\label{spinorscalar}
  \conj{\psi} \cdot \psi
  =
  \psi^\dagger \bGamma \psi
  \ ,
\end{equation}
which is real given that the eigenvalues of $\bGamma$ are real.
In zero time dimensions the scalar product of a spinor
with its conjugate is always positive,
but with one or more time dimensions the scalar product of a spinor
with its conjugate can be either positive or negative (or zero).

The scalar product $\conj{\psi} \cdot \bGamma \chi$ is
\begin{equation}
  \conj{\psi} \cdot \bGamma \chi
  =
  \psi^\dagger \chi
  \ .
\end{equation}
In particular,
$\conj{\psi} \cdot \bGamma \psi$ is real and positive,
\begin{equation}
\label{conjpsipsiKM}
  \conj{\psi} \cdot \bGamma \psi
  =
  \psi^\dagger \psi
  \ .
\end{equation}

\subsection{Conjugate multivectors}

The complex conjugate (with respect to $\im$)
$\ba^\ast$ of a multivector $\ba = a^A \bgamma_A$
is defined to be its complex conjugate in the
chiral representation~(\ref{bepsilontrix}) of multivectors,
\begin{equation}
  \ba^\ast
  =
  a^{A \ast} \bgamma_A^\ast
  \ .
\end{equation}
In the representation~(\ref{bepsilontrix}),
the chiral basis vectors $\bgamma_k$ and $\bgamma_{\bar  k}$
(and the final vector $\bgamma_N$ if $N$ is odd) are real,
so the orthonormal basis vectors $\bgamma_k^+$ and $\bgamma_k^-$
are respectively real and imaginary.
The conjugate $\conj{\ba}$ of a multivector $\ba = a^A \bgamma_A$ is
defined to be,
consistent with the definition~(\ref{conjspinor})
of the conjugate of a spinor,
\begin{equation}
  \conj{\ba}
  \equiv
  \Cp \ba^\ast \Cp^{-1}
  \ .
\end{equation}
The conjugate multivector $\conj{\ba}$
rotates under a rotor $R$ in the same way as the multivector $\ba$,
\begin{equation}
  R : \ 
  \conj{\ba}
  \equiv
  \Cp \ba^\ast \Cp^{-1}
  \rightarrow
  \Cp ( R \ba \reverse{R} )^\ast \Cp^{-1}
  =
  \Cp R^\ast \ba^\ast \reverse{R}^\ast \Cp^{-1}
  =
  R \Cp \ba^\ast \Cp^{-1} \reverse{R}
  =
  R \conj{\ba} \reverse{R}
  \ .
\end{equation}
Conjugation is multiplicative over multivectors,
and over multivectors with spinors,
\begin{equation}
  \longconj{\ba \bb} = \conj{\ba} \conj{\bb}
  \ , \quad
  \longconj{\ba \psi} = \conj{\ba} \conj{\psi}
  \ .
\end{equation}


The conjugate $\conj{\ba}$ of an outer product
$\ba \equiv \psi \conj{\chi} \spinordot$
of spinors is related to the outer product
$\conj{\psi} \chi \spinordot$
by
\begin{equation}
\label{conjspinorouterproduct}
  \ba
  \equiv
  \psi \conj{\chi} \spinordot
  \ , \quad
  \conj{\ba}
  \equiv
  \longconj{\psi \conj{\chi} \spinordot}
  =
  \pm
  \conj{\psi} \chi \spinordot
  \ ,
\end{equation}
where the $\pm$ sign is the symmetry of the spinor metric,
Table~\ref{spinormetricsymmetrytab}.

The conjugate of a basis multivector $\bgamma_A$ is defined to be
\begin{equation}
\label{conjbasisvectorN}
  \conj{\bgamma}_A
  \equiv
  \Cp \bgamma_A^\ast \Cp^{-1}
  \ ,
\end{equation}
so that a conjugate multivector $\conj{\ba}$ is
\begin{equation}
\label{conjbasisvectorN}
  \conj{\ba}
  =
  a^{A\ast} \conj{\bgamma}_A
  \ .
\end{equation}
The conjugate of a spacelike or timelike orthonormal basis vector $\bgamma_m$ is
(if $\bgamma_m$ is a timelike basis vector,
$\bgamma_m$ here means the timelike vector itself,
whose square is $-1$)
\begin{equation}
\label{conjbasisvectoraN}
  \conj{\bgamma}_m
  =
  \pm (-)^M
  \bgamma_m
  \ ,
\end{equation}
where the $\pm$ factor is as given in Table~\ref{spinormetricgammaatab}.

A real subalgebra of the complex geometric algebra may be obtained
by restricting to multivectors
satisfying the reality condition that they are their own conjugates,
\begin{equation}
\label{conjaaKM}
  \conj{\ba}
  =
  \ba
  \ .
\end{equation}
Conjugates of orthonormal basis vectors are plus or minus themselves
per equation~(\ref{conjbasisvectoraN}).
If the overall sign $\pm (-)^M$ in equation~(\ref{conjbasisvectoraN}) is $+$,
as it is for example in the (3+1)-dimensional Dirac algebra,
then the real subalgebra consists of real linear combinations
of orthonormal basis multivectors.
If the sign $\pm (-)^M$ in equation~(\ref{conjbasisvectoraN}) is $-$,
then the real subalgebra consists of linear combinations of
odd-grade orthonormal multivectors with pure imaginary coefficients
and even-grade orthonormal multivectors with pure real coefficients.

\section{Conclusions}
\label{conclusions-sec}

All matter (fermions) is made from spinors.
All forces arise from symmetries of spinors.
It would not be surprising to find that spinors are
at the heart of a theory that unifies all forces of nature.

The index of a spinor in $N$ dimensions is a bitcode with $[N/2]$ bits.
It is delightful to contemplate that the laws of physics
might be written in a language whose letters are a spinor bitcode.

Whereas a vector sticks out in one dimension at a time,
a spinor sticks out in all dimensions at once.
A basis spinor is either up ($\uparrow$) or down $(\downarrow)$,
not zero, in each pair of dimensions.

While it is common to consider the geometric (Clifford) algebra
as the foundation upon which spinors are built,
this paper advances the alternative view that spinors are more fundamental
than the geometric algebra.
To this end,
the paper introduces the supergeometric algebra,
which consists of complex linear combinations of scalars (complex numbers),
column spinors, row spinors, and multivectors,
and the various multiplication rules that connect them,
equations~(\ref{spinormultiplication}).
The exclusion principle, which is a central property of spinors in physics,
is not apparent in the geometric algebra,
but becomes apparent in the language of the supergeometric algebra.

Among other things, this paper gives a proof,
section~\ref{algebraouterproduct-sec},
of the Brauer-Weyl (1935) \cite{Brauer:1935} theorem,
that the algebra of outer products of spinors
is isomorphic to the geometric algebra of multivectors.
The theorem formalizes the notion that
the geometric algebra is the `square' of the supergeometric algebra,
or equivalently that
the supergeometric algebra is the `square root' of geometry
\cite{Trautman:1979}.


The algebra of outer products of spinors yields not only vectors
but the entire geometric algebra of multivectors.
Since multivectors of grade $p$ are associated with geometric objects
of dimension $p$,
it is intriguing to infer that geometric objects of arbitrary dimension,
which are called branes in string theory ($M$ theory),
are an intrinsic part of a theory built from spinors.

This paper introduces the convenient notation
$\psi \spinordot$ with a trailing dot to signify a row spinor,
equation~(\ref{rowspinor}).
The trailing dot represents
the spinor metric tensor $\varepsilon$.
A row spinor $\psi \spinordot$ times a column spinor $\chi$
defines their inner product $\psi \cdot \chi$, a scalar.
A column spinor $\chi$ times a row spinor $\psi \spinordot$
defines their outer product $\chi \psi \spinordot$, which transforms
like a multivector.
The trailing-dot notation makes transparent the application
of the associative rule to a sequence of products of row and column spinors,
equation~(\ref{spinorassociative}).

\subsection*{Data statement}
Data sharing is not applicable to this article as no new data were created or analyzed in this study.

\subsection*{Acknowledgment}
This research was supported in part by FQXI mini-grant FQXI-MGB-1626.

\appendix

\section{The Pauli supergeometric algebra}
\label{spinorgeom3D-app}

The supergeometric algebra in 2 or 3 spatial dimensions
consists of Pauli spinors and the Pauli algebra,
and their various products.

In $N = 2$ or 3 dimensions,
there is just $[N/2] = 1$ bit,
and $2^{[N/2]} = 2$ basis column spinors
with respectively spin up and spin down,
\begin{equation}
  \bepsilon_a = \{ \bepsilon_{\uparrow} , \bepsilon_{\downarrow} \}
  \ .
\end{equation}
As column vectors in the chiral representation, the two basis spinors are
\begin{equation}
  \bepsilon_{\uparrow}
  =
  \left(
  \begin{array}{c}
  1 \\ 0
  \end{array}
  \right)
  \ , \quad
  \bepsilon_{\downarrow}
  =
  \left(
  \begin{array}{c}
  0 \\ 1
  \end{array}
  \right)
  \ .
\end{equation}
In Dirac bra-ket notation, the spinors would be written
$\bepsilon_{\uparrow} = \lvert \uparrow \rangle$
and
$\bepsilon_{\downarrow} = \lvert \downarrow \rangle$.
The outer products of the 2 column basis spinors
$\bepsilon_a$
with the 2 row basis spinors
$\bepsilon_b \spinordot$
form 4 outer products,
yielding the 2D geometric algebra with $2^2 = 4$ basis elements.
The 3D geometric algebra,
the Pauli algebra,
has $2^3 = 8$ basis elements,
but is projected into 2D by identifying the pseudoscalar
$I_3$ with $\im$ times the unit matrix,
equation~(\ref{IiNodd}).

The orthonormal vectors $\bgamma_a$
of the Pauli algebra are commonly denoted
$\sigma_a = \{ \sigma_1 , \sigma_2 , \sigma_3 \}$.
In the notation of paired orthonormal vectors $\bgamma^\pm_k$
of the present paper,
the orthonormal Pauli vectors are, equations~(\ref{gammavecNrep}),
\begin{equation}
\label{Pauli}
  \bgamma^+_1
  =
  \sigma_1 \equiv
  \left(
    \begin{array}{cc}
      0 & 1 \\
      1 & 0
    \end{array}
  \right)
  \, , \ 
  \bgamma^-_1
  =
  \sigma_2 \equiv
  \left(
    \begin{array}{cc}
      0 & -\im \\
      \im & 0
    \end{array}
  \right)
  \, , \ 
  \bgamma_3
  =
  \sigma_3 \equiv
  \left(
    \begin{array}{cc}
      1 & 0 \\
      0 & -1
    \end{array}
  \right)
  \, .
\end{equation}
The 2D and 3D pseudoscalars $I_2$ and $I_3$ are
\begin{equation}
  I_2
  \equiv
  \bgamma^+_1 \bgamma^-_1
  =
  \im
  \left(
    \begin{array}{cc}
      1 & 0 \\
      0 & -1
    \end{array}
  \right)
  \ , \quad
  I_3
  \equiv
  \bgamma^+_1 \bgamma^-_1 \bgamma_3
  =
  \im
  \left(
    \begin{array}{cc}
      1 & 0 \\
      0 & 1
    \end{array}
  \right)
  \ .
\end{equation}
The third vector $\bgamma_3$ is, modulo a phase,
equal to the 2-dimensional pseudoscalar,
equation~(\ref{gammaNodd}),
$\bgamma_3 = -\im I_2$.
The chiral basis vectors~(\ref{gammaapm}) of the Pauli algebra are
\begin{equation}
\label{gammap}
  \bgamma_1
  =
  {\sigma_1 + \im \sigma_2 \over \sqrt{2}}
  =
  \left(
  \begin{array}{cc}
     0 & \sqrt{2} \\
     0 & 0
  \end{array}
  \right)
  \ , \quad
  \bgamma_{\bar 1}
  =
  {\sigma_1 - \im \sigma_2 \over \sqrt{2}}
  =
  \left(
  \begin{array}{cc}
     0 & 0 \\
     \sqrt{2} & 0
  \end{array}
  \right)
  \ .
\end{equation}
The bivectors of the 3D Pauli algebra are the pseudovectors
$I_3 \sigma_a = \im \sigma_a$.

The algebra of outer products of spinors in 2 or 3 dimensions is as follows.
The antisymmetric outer products of spinors
form a scalar singlet,
\begin{equation}
\label{singlet}
  [ \bepsilon_{\downarrow} , \bepsilon_{\uparrow} ] \spinordot
  =
  1
  \ ,
\end{equation}
where the $1$ on the right hand side denotes the unit element,
the $2 \times 2$ identity matrix.
The trailing dot on the commutator
indicates that the right partner of each product is a row spinor,
$[ \bepsilon_{\uparrow} , \bepsilon_{\downarrow} ] \spinordot = \bepsilon_{\uparrow} \bepsilon_{\downarrow} \spinordot - \bepsilon_{\downarrow} \bepsilon_{\uparrow} \spinordot$.
The spin charge of the singlet~(\ref{singlet}) is zero
according to the rule~(\ref{kchargespinor}).
The spin charge, zero, of the left and right hand sides match, as they should.

The symmetric outer products of spinors
yield the chiral vectors of the Pauli algebra,
\begin{equation}
\label{triplet}
  \{ \bepsilon_{\uparrow} , \bepsilon_{\uparrow} \} \spinordot
  =
  \sqrt{2} \,
  \bgamma_1
  \ , \quad
  \{ \bepsilon_{\uparrow} , \bepsilon_{\downarrow} \} \spinordot
  =
  - \, \bgamma_3
  \ , \quad
  \{ \bepsilon_{\downarrow} , \bepsilon_{\downarrow} \} \spinordot
  =
  -
  \sqrt{2} \,
  \bgamma_{\bar 1}
  \ .
\end{equation}
The spin charges of the vectors~(\ref{triplet}) are respectively
$+1$, $0$, $-1$
according to the rules~(\ref{kcharge}) and (\ref{kchargespinor}).
The spin charges of the left and right hand sides match, as they should.

The spinor metric $\varepsilon$ in 3 dimensions is,
equations~(\ref{pspinormetricp}) or~(\ref{altpspinormetricp}),
\begin{equation}
\label{egamma2}
  \varepsilon
  \equiv
  \im \bgamma^-_1
  =
  \left(
  \begin{array}{cc}
  0 & 1 \\
  -1 & 0
  \end{array}
  \right)
  \ .
\end{equation}
The spinor metric $\varepsilon$ is antisymmetric,
in agreement with Table~\ref{spinormetricsymmetrytab}.
Despite the equality of $\varepsilon$ and $\im \bgamma^-_1$
in the Pauli representation,
$\varepsilon$ is defined to transform as a spinor tensor
under spatial rotations,
not as an element of the geometric algebra.
As always in a geometric algebra with only spatial (no time) dimensions,
the conjugation operator $\Cp$ coincides with the spinor metric $\varepsilon$,
equation~(\ref{arbdimensionCp}),
\begin{equation}
\label{CPauli}
  \Cp
  =
  \varepsilon
  \ .
\end{equation}

\section{The Dirac supergeometric algebra}
\label{spinorgeom31D-app}

The supergeometric algebra in $N = 3{+}1$ spacetime dimensions
consists of Dirac spinors and the Dirac algebra, and their various products.
The Dirac algebra is summarized in a notation consistent with the present paper
in Appendix~A of \cite{Hamilton:2022b},
and that summary will not be repeated here.
The present Appendix gives the relation between
outer products of basis Dirac spinors
and the basis multivectors of the Dirac algebra,
equations~(\ref{singletR})--(\ref{quadrupletpseudodirac}).

The Dirac algebra has $[N/2] = 2$ bits,
a boost bit $\Uparrow$ or $\Downarrow$,
and a spin bit $\uparrow$ or $\downarrow$.
There are $2^{[N/2]} = 4$ basis spinors,
which group into two massless Weyl spinors of opposite chirality,
right- and left-handed.
A Dirac (Weyl) spinor is right-handed if its boost and spin bits align,
left-handed if its bits antialign.
The 4 Dirac basis spinors,
ordered with the right-handed pair first, then the left-handed pair, are
\begin{equation}
  \bepsilon_a
  =
  \{ \bepsilon_{\Upup} , \bepsilon_{\Downdown} , \bepsilon_{\Downup} , \bepsilon_{\Updown} \}
  \ .
\end{equation}

Orthonormal basis vectors in the Dirac algebra are conventionally denoted
$\bgamma_m$, $m = 0,1,2,3$.
In the notation of paired orthonormal vectors $\bgamma^\pm_k$
of the present paper, the traditional basis vectors are
\begin{equation}
\label{gammadiracpm}
  \{ \bgamma_1 , \, \bgamma_2 , \, \bgamma_3 , \, \bgamma_0 \}
  =
  \{ \bgamma_1^+ , \, \bgamma_1^- , \, \bgamma_2^+ , \, \im \bgamma_2^- \}
  \ .
\end{equation}
A common notation for chiral vectors in the Newman-Penrose community is,
in terms of orthonormal vectors,
$\bgamma_v \equiv {( \bgamma_0 + \bgamma_3 ) / \sqrt{2}}$,
$\bgamma_u \equiv {( \bgamma_0 - \bgamma_3 ) / \sqrt{2}}$,
$\bgamma_+ \equiv {( \bgamma_1 + \im \bgamma_2 ) / \sqrt{2}}$,
$\bgamma_- \equiv {( \bgamma_1 - \im \bgamma_2 ) / \sqrt{2}}$.
In the notation $\bgamma_k$ and $\bgamma_{\bar  k}$
of paired chiral vectors
used elsewhere in this paper, equations~(\ref{gammaapm}),
the chiral (Newman-Penrose) basis vectors are
\begin{equation}
  \{ \bgamma_+ , \, \bgamma_- , \, \bgamma_v , \, \bgamma_u \}
  =
  \{ \bgamma_1 , \, \bgamma_{\bar 1} , \, \bgamma_2 , \, -\bgamma_{\bar 2} \}
  \ .
\end{equation}

The 8 outer products of basis spinors of like chirality
map to even multivectors of the spacetime algebra as follows.
The boost and spin charges of the left and right hand sides
of each of equations~(\ref{singletR})--(\ref{bivectorL})
below match, as they should.
The antisymmetric outer products of right-handed spinors
form a right-handed scalar singlet,
\begin{equation}
\label{singletR}
  [ \bepsilon_{\Downdown} , \bepsilon_{\Upup} ] \spinordot
  =
  \tfrac{1}{2} ( 1 + \varkappa_4 )
  \ ,
\end{equation}
where $\varkappa_4$ is the chiral operator in 4 dimensions,
equation~(\ref{kappaN}),
commonly denoted $\gamma_5$ in the traditional Dirac community.
The trailing dot on the commutator
indicates that the right partner of each product is a row spinor,
$[ \bepsilon_{\Downdown} , \bepsilon_{\Upup} ] \spinordot = \bepsilon_{\Downdown} \bepsilon_{\Upup} \spinordot - \bepsilon_{\Upup} \bepsilon_{\Downdown} \spinordot$.
Similarly the antisymmetric outer products of left-handed spinors
form a left-handed scalar singlet,
\begin{equation}
\label{singletL}
  [ \bepsilon_{\Updown} , \bepsilon_{\Downup} ] \spinordot
  =
  \tfrac{1}{2} ( 1 - \varkappa_4 )
  \ .
\end{equation}
The symmetric outer products of right-handed spinors
form the three right-handed bivectors,
\begin{equation}
\label{bivectorR}
  \{ \bepsilon_{\Upup} , \bepsilon_{\Upup} \} \spinordot
  =
  -
  \bgamma_{12}
  \ , \quad
  \{ \bepsilon_{\Upup} , \bepsilon_{\Downdown} \} \spinordot
  =
  -
  \tfrac{1}{2} ( \bgamma_{1{\bar 1}} + \bgamma_{2{\bar 2}} )
  \ , \quad
  \{ \bepsilon_{\Downdown} , \bepsilon_{\Downdown} \} \spinordot
  =
  -
  \bgamma_{{\bar 1}{\bar 2}}
  \ ,
\end{equation}
while the symmetric outer products of left-handed spinors
form the three left-handed bivectors,
\begin{equation}
\label{bivectorL}
  \{ \bepsilon_{\Downup} , \bepsilon_{\Downup} \} \spinordot
  =
  -
  \bgamma_{1{\bar 2}}
  \ , \quad
  \{ \bepsilon_{\Downup} , \bepsilon_{\Updown} \} \spinordot
  =
  -
  \tfrac{1}{2} ( \bgamma_{1{\bar 1}} - \bgamma_{2{\bar 2}} )
  \ , \quad
  \{ \bepsilon_{\Updown} , \bepsilon_{\Updown} \} \spinordot
  =
  -
  \bgamma_{{\bar 1}2}
  \ .
\end{equation}

The 8 outer products of basis spinors of opposite chirality
map to odd multivectors of the spacetime algebra as follows.
Again, the boost and spin charges of the left and right hand sides of each of
equations~(\ref{quadrupletdirac})--(\ref{quadrupletpseudodirac})
below match, as they should.
The 4 symmetric outer products of right- with left-handed spinors
yield the 4 chiral basis vectors,
\begin{subequations}
\label{quadrupletdirac}
\begin{alignat}{3}
  \{ \bepsilon_{\Upup} , \bepsilon_{\Downup} \} \spinordot
  &=
  \tfrac{1}{\sqrt{2}}
  \bgamma_1
  \ , \quad
  &\{ \bepsilon_{\Downdown} , \bepsilon_{\Updown} \} \spinordot
  &=
  -
  \tfrac{1}{\sqrt{2}}
  \bgamma_{\bar 1}
  \ ,
\\
  \{ \bepsilon_{\Upup} , \bepsilon_{\Updown} \} \spinordot
  &=
  -
  \tfrac{1}{\sqrt{2}}
  \bgamma_2
  \ , \quad
  &\{ \bepsilon_{\Downdown} , \bepsilon_{\Downup} \} \spinordot
  &=
  -\tfrac{1}{\sqrt{2}}
  \bgamma_{\bar 2}
  \ ,
\end{alignat}
\end{subequations}
while the 4 antisymmetric outer products of right- with left-handed spinors
yield the 4 chiral basis pseudovectors,
\begin{subequations}
\label{quadrupletpseudodirac}
\begin{alignat}{3}
  [ \bepsilon_{\Upup} , \bepsilon_{\Downup} ] \spinordot
  &=
  \tfrac{1}{\sqrt{2}}
  \varkappa_4
  \bgamma_1
  \ , \quad
  &[ \bepsilon_{\Downdown} , \bepsilon_{\Updown} ] \spinordot
  &=
  -
  \tfrac{1}{\sqrt{2}}
  \varkappa_4
  \bgamma_{\bar 1}
  \ ,
\\
  [ \bepsilon_{\Upup} , \bepsilon_{\Updown} ] \spinordot
  &=
  -
  \tfrac{1}{\sqrt{2}}
  \varkappa_4
  \bgamma_2
  \ , \quad
  &[ \bepsilon_{\Downdown} , \bepsilon_{\Downup} ] \spinordot
  &=
  -
  \tfrac{1}{\sqrt{2}}
  \varkappa_4
  \bgamma_{\bar 2}
  \ .
\end{alignat}
\end{subequations}

\section{Other options for the spinor metric tensor}
\label{spinormetric-app}

If $N$ is odd,
and if the odd algebra is constructed
as described in section~\ref{oddN2-sec}
by embedding the odd algebra in one extra dimension and treating
either the final (odd) dimension $\bgamma_N$ or
the extra (even) dimension $\bgamma_{N+1}$ as a scalar,
then there are potentially further options for the spinor metric
beyond those described in section~\ref{spinormetric-sec}.

The invariance condition~(\ref{pspinormetrictensorN})
need hold only for rotors not involving the scalar dimension
$\bgamma_N$ or $\bgamma_{N+1}$.
If the scalar dimension is the odd dimension $\bgamma_N$,
then $\bgamma_N$ can be dropped from the standard spinor metric $\varepsilon$,
leaving $\varepsilon$ in $N{-}1$ dimensions.
If the scalar dimension is the even dimension $\bgamma_{N+1}$,
then $\im \bgamma_{N+1}$ can be adjoined to
the alternative spinor metric $\varepsilon_{\rm alt}$,
giving $\varepsilon_{\rm alt}$ in $N{+}1$ dimensions.
The resulting spinor metrics, distinguished with a prime, are
\begin{equation}
\label{pspinormetricpodd2}
  \varepsilon^\prime_N
  =
  \varepsilon_N
  \bgamma_N
  =
  \varepsilon_{N-1}
  \ , \quad
  \varepsilon^\prime_{{\rm alt} , N}
  =
  \varepsilon_{{\rm alt} , N}
  \im
  \bgamma_{N+1}
  =
  \varepsilon_{{\rm alt} , N+1}
  \quad
  \mbox{($N$ odd)}
  \ .
\end{equation}

\printbibliography

\end{document}